\documentclass[letterpaper]{JHEP3}
\usepackage{epsfig,verbatim}

\newcommand{\eref}[1]{(\ref{#1})}

\renewcommand\({\left(}
\renewcommand\){\right)}
\renewcommand\[{\left[}
\renewcommand\]{\right]}

\def\be{\begin{equation}}
\def\ee{\end{equation}}
\def\bea{\begin{eqnarray}}
\def\eea{\end{eqnarray}}

\providecommand{\V}{\mathcal{V}}

\def\sss{\scriptscriptstyle}

\def\d{\exd}

\def\ba{\begin{eqnarray}}
\def\ea{\end{eqnarray}}
\def\be{\begin{equation}}
\def\ee{\end{equation}}

\def\A{\mathcal{A}}

\def\F{\mathcal{F}}

\def\V{\mathcal{V}}

\def\nn{\nonumber}

\def\d{\mathrm{d}}

\def\({\left(}
\def\){\right)}

\def\pref#1{(\ref{#1})}

\newcommand{\roughly}[1]{\mathrel{\raise.3ex\hbox{$#1$\kern-0.85em
\lower1ex\hbox{$\sim$}}}}

\newcommand{\lsim}{\roughly<}
\newcommand{\gsim}{\roughly>}

\title{Axionic $D3$-$D7$ Inflation}

\author{C.P. Burgess,${}^{1-3}$ J.M. Cline${}^4$ and M. Postma${}^5$ \\
${}^1$ Perimeter Institute for Theoretical Physics, Waterloo ON,
N2L 2Y5, Canada.\\
${}^2$
Physics \& Astronomy, McMaster University, Hamilton ON, L8S 4M1,
Canada. \\
${}^3$ Theory Division, CERN, CH-1211 Geneva 23, Switzerland. \\
${}^4$ Physics Dept., McGill University, Montr\'eal, QC, H3A 2T8,
Canada.\\
${}^5$ NIKHEF, Kruislaan 409, 1098 Amsterdam, The Netherlands. }

\date{}
\abstract {We study the motion of a $D3$ brane moving within a Type
IIB string vacuum compactified to 4D on $K3 \times T_2/Z_2$ in the
presence of $D7$ and $O7$ planes. We work within the effective 4D
supergravity describing how the mobile $D3$ interacts with the
lightest bulk moduli of the compactification, including the effects of
modulus-stabilizing fluxes. We seek inflationary solutions to the
resulting equations, performing our search numerically in order to
avoid resorting to approximate parameterizations of the low-energy
potential. We consider uplifting from $D$-terms and from the
supersymmetry-breaking effects of anti-$D3$ branes. We find examples
of slow-roll inflation (with anti-brane uplifting) with the mobile
$D3$ moving along the toroidal directions, falling towards a $D7$-$O7$
stack starting from the antipodal point. The inflaton turns out to be
a linear combination of the brane position and the axionic partner of
the $K3$ volume modulus, and the similarity of the potential along the
inflaton direction with that of racetrack inflation leads to the
prediction $n_s \le 0.95$ for the spectral index. The slow roll is
insensitive to most of the features of the effective superpotential,
and requires a one-in-$10^4$ tuning to ensure that the torus is close
to square in shape. We also consider $D$-term inflation with the $D3$
close to the attractive $D7$, but find that for a broad (but not
exhaustive) class of parameters the conditions for slow roll tend to
destabilize the bulk moduli. In contrast to the axionic case, the best
inflationary example of this kind requires the delicate adjustment of
potential parameters (much more than the part-per-mille level), and
gives inflation only at an inflection point of the potential (and so
suffers from additional fine-tuning of initial conditions to avoid an
overshoot problem).

}

\preprint{NIKHEF 2008-031}

\begin{document}

%%%%%%%%%%%%%%%%%%%%%%%%%%%%%%%%%%%%%%%%%%%%%%%%%%%%%%%%%%%%%%
%%%%%%%%%%%%%%%%%%%%%%%%%%%%%%%%%%%%%%%%%%%%%%%%%%%%%%%%%%%%%%

\section{Introduction}

The advent of tools for fixing moduli in string theory has opened
up the possibility for surveying where slow-roll inflation occurs
among string vacua, with the result (so far) that it appears to be
relatively rare, but not impossible. This survey has revealed a
variety of potential inflationary mechanisms, with the inflaton
residing either among open or closed string modes \cite{revs}.

Among the most interesting of these mechanisms is that of $D3$-$D7$
inflation \cite{d3d7}, for which the inflaton is the separation
between mobile $D3$ branes as they approach static stacks of $D7$
branes. Besides sharing many of the attractive features of
brane-antibrane models \cite{DT,BBar,BBar2,KKLMMT}, this scenario
potentially has the additional advantage that the final $D3$-$D7$
collision may be better understood, with the possibility of the
$D3$ dissolving into the $D7$ to leave a supersymmetric state.
Furthermore, stacks of coincident $D7$ and $O7$ planes can source
flat transverse geometries and constant dilaton configurations,
among which are the well-studied compactifications on $K3 \times
T_2/Z_2$. One might expect the prospects for finding slow roll for
$D3$ motion for such geometries to be better than for a generic
Calabi-Yau.

Additional progress became possible with the application of Type
IIB modulus-stabiliza-tion techniques \cite{GKP,KKLT} to $K3
\times T_2/Z_2$ geometries \cite{tt,ak,GKTT}. This opened up the
possibility of understanding the low energy dynamics within the
framework of the effective 4D supergravity, with all the
additional control over the calculation that this brings. Until
recently one ingredient remained missing for performing a more
systematic 4D study of $D3$ motion in these systems, and this was
the 4D supergravity formulation of the forces acting on a mobile
$D3$ brane once supersymmetry becomes broken (such as by the
addition of magnetic fluxes in the 7-brane world volume). This
missing step was removed with the analysis
\cite{bhk,baumann,uplifting} of how the $D3$ back-reacts on the
$D7$ geometry, and thereby introduces a dependence on the $D3$
position into the energetics of $D7$ physics (like gaugino
condensation or magnetic fluxes).

A first study of $D3$ motion in $K3 \times T_2/Z_2$ was recently
performed in ref.~\cite{haack}, who also made a preliminary search
for inflationary solutions using a semi-phenomenological
potential. This potential was meant to parameterize the important
features of the low-energy supergravity in the limit when the $D3$
and $D7$ are in close proximity. In particular, it includes a
combination of a logarithmic `Coleman-Weinberg' (CW) potential
describing the attraction of a $D3$ towards a $D7$ on which
supersymmetry has been broken by fluxes, the $D$-term energy
generated by this flux \cite{bkq}, plus the nonperturbative
superpotential generated by gaugino condensation on a $D7$ stack
located at a different fixed point. Their search identified a
putative slow-roll inflationary regime when the mobile $D3$
approaches very closely one of the $D7/O7$ stacks.

In this note we extend their analysis in several ways.
\begin{itemize}
\item First we follow the evolution of more of the twenty-odd bulk
moduli of $K3 \times T_2/Z_2$. After describing the low-energy
supergravity in some generality, we follow the dynamics of two of
these complex moduli in addition to that of the $D3$-brane
position. We do so because it is only when at least two of the
bulk moduli are kept that the full no-scale form of the leading
low-energy potential is manifest.
\item Second, we search numerically for inflationary solutions,
allowing the use of the actual $F$- and $D$-term potentials of the
low-energy supergravity, rather than an approximate
semi-phenomenological potential. Since we need not rely on
expansions in the $D3$-$D7$ distance, we can both test the domain
of validity of the approximate forms used by earlier workers, and
can search for inflation when the $D3$ is far from the $D7$.
\item Third, we consider two types of `uplifting' physics,
required to assure the potential is minimized at a Minkowski
vacuum after inflation ends. Following \cite{haack} we examine
$D$-term uplifting as generated by $D7$ fluxes. But due to present
difficulties in obtaining these from explicit string vacua on $K3
\times T_2/Z_2$ we also explore uplifting due to anti-$D3$ branes
{\it \`a la} KKLT \cite{KKLT}.
\end{itemize}

Our search reveals several examples of slow-roll inflation, in all
cases requiring some degree of tuning of the parameters of the
potential. We focus on inflationary trajectories with the $D3$ moving
along the torus. This is because the K\"ahler potential has a shift
symmetry in the torus coordinate which may protect that direction from
getting large corrections from the non-perturbative $F$-term
potential. Our best example occurs when the $D3$ falls between two
stacks of $D7$'s, due to forces ultimately driven by nonperturbative
physics (like gaugino condensation or Euclidean $D$-branes) occurring
on yet a third such stack. Inflation occurs when the $D3$ starts at
the antipodal point, within the torus, of the $D7$'s on which the
nonperturbative physics occurs. In this case the tuning required is
quite mild, with the inflationary roll largely insensitive to other
parameters once the torus is adjusted to be close to square. The
inflaton direction turns out to be a combination of the $D3$ position
and the axionic partner of the $K3$ volume modulus, leading to a
situation similar to the racetrack inflation model \cite{racetrack}.
%, but with less tuning of parameters required. 
Since the starting position is at a local
maximum of the inflaton direction, eternal topological inflation
can remove the need for explaining the initial conditions.
Uplifting is provided in this example by the presence of an
anti-$D3$ brane.

We also search for inflation in the regime of ref.~\cite{haack},
where the $D3$ is close to a stack of $D7$'s on which
supersymmetry-breaking fluxes provide the inflationary energy
density. In this case we find inflation much more difficult to
achieve, largely because we are unable to realize the parameter
choices required for their slow roll within our 4D supergravity.
We {\it are} able to obtain slow-roll inflation in this regime,
however, although only by using a delicately tuned (to within a
part per million) choice of potential parameters. What is
troublesome, however, is that the inflationary regime that results
arises near an inflection point of the potential, rather than a
local maximum. This has the disadvantage of requiring a several
percent tuning of the initial conditions to avoid having an
overshoot problem.

Our discussion is organized as follows.  In section 2 we review
the underlying theoretical ingredients leading to the low-energy
effective action for the inflaton.  In section 3 we develop the
Lagrangian explicitly, in terms of the F-term and D-term
contributions and possible uplifting by anti-D3 branes in both
warped and unwarped backgrounds, but restricted to the fields
whose dynamics we wish to follow. Section 4 describes the two
examples of inflation described above: the racetrack-like model
starting from the antipodal point of the attracting $D7$ brane
(using $\overline{D3}$ uplifting), and the inflection point model
(with $D$-term uplifting) where the $D3$ is near the $D7$. We
present our conclusions in section 5. The appendix contains
results concerning the no-scale property of the K\"ahler
potential, our conventions for Jacobi theta functions, and scaling
properties of the potential under certain re-scalings of the
Lagrangian parameters.

%%%%%%%%%%%%%%%%%%%%%%%%%%%%%%%%%%%%%%%%%%%%%%%%%%%%%%%%%%%%%%

\section{Low Energy Dynamics on $K3 \times T_2/Z_2$}

In this section we develop the general properties of the 4D
supergravity describing the low-energy behaviour of $K3 \times
T_2/Z_2$, before specializing in the next section to the moduli
playing a direct role in the inflationary scenario.

\subsection{The field content}
\label{sec:fieldcontent}

Our starting point is a Type IIB string vacuum compactified on $K3
\times T_2/Z_2$, in the presence of moduli-stabilizing 3-form
fluxes \cite{GKP,KKLT}, such as studied by \cite{tt,ak}. The
orientifold $Z_2$ acts on the torus by reflecting its (complex)
coordinates, $z \to -z$, leading to $O7$-planes located at four
fixed points. Taking the torus to be defined by the parallelogram
$z \simeq z + 1$ and $z \simeq z + \tau$, with $\tau$ the complex
modulus satisfying Im $\tau > 0$, these fixed points are situated
at $z = 0$, $\frac12$, $\frac12\tau$ and $\frac12(1+\tau)$. The
$D7$ tadpole conditions are satisfied when each $O7$ plane is
accompanied by 4 $D7$'s, all wrapping the $K3$. If the 4 $D7$'s are
coincident with the corresponding $O7$, they do not source the
dilaton field, which can therefore remain constant along the
toroidal directions. The $D3$ tadpole condition requires the
number of $D3$ branes plus a flux integral to sum to 24 \cite{tt}.

$K3$ is a Ricci flat space having two complex dimensions which
naturally arises in supersymmetric compactifications of string
theory to 4D \cite{strominger}, being in many ways the
lower-dimensional analog of the three (complex) dimensional Calabi
Yau spaces. It has a very rich topology \cite{K3}, with Hodge
numbers $h_{10} = h_{01} = 0$, $h_{00} = h_{20} = h_{02} = h_{22}
= 1$ and $h_{11} = 20$, leading to an Euler number $\chi = 24$.
These are the same as for the orbifold $T_4/Z_2$, say, whose 16
fixed points can be regarded as the degenerate limit of 16 of the
22 nontrivial 2-cycles on $K3$.

For Type IIB string compactifications this topology leads to
low-energy moduli, $T^\alpha = \xi^\alpha + i \beta^\alpha$. Some
of these moduli are stabilized (at leading order in the $\alpha'$
and string loop expansions) once the 3-form fluxes are turned on,
and these fluxes can preserve zero, one or two low-energy 4D
supersymmetries \cite{tt}. The rest of the moduli can be
stabilized in principle by nonperturbative effects \cite{ak}. The
dynamics of this stabilization can be described by a low-energy 4D
supergravity provided that the supersymmetry breaking scales are
kept parametrically small compared with the Kaluza-Klein (KK)
scale, as we assume to be the case in what follows.

To this geometry we imagine adding one or more of the following
optional features.
\begin{itemize}
\item For inflationary purposes, we imagine adjusting the fluxes
to allow the presence of a mobile $D3$ brane situated at a point
in the extra dimensions. We argue below that the physics that
stabilizes the various K\"ahler moduli on $K3$ tends also to
stabilize the motion of this brane in the $K3$ directions,
although it can be relatively free to move along the toroidal
directions, with complex coordinate $z$.
\item It is often useful to entertain the presence of an anti-$D3$
brane, in order to uplift the minimum of the potential to zero.
Ultimately, the necessity for doing so reflects our poor
understanding of the cosmological constant problem, and we regard
such an anti-brane to represent a parametrization of whatever
mechanism properly solves this problem in the string vacuum of
interest. When doing so it is often useful to sequester the
antibrane into a warped throat on $K3 \times T_2/Z_2$, such as was
studied in ref.~\cite{dasgupta}. This has several advantages.
Besides helping to localize the $\overline{D3}$, which reduces its
energy by sitting in the throat, it also reduces its impact on the
dynamics of the mobile $D3$ brane, by suppressing their direct
`Coulomb' attraction.
\item It is also possible to add background magnetic 2-form
fluxes, ${\cal F}$, for gauge fields residing on the D7 branes, in
order to uplift the potential at its minimum \cite{bkq}. If such
fluxes are present they typically gauge some of the axion
symmetries under which the imaginary parts of the moduli shift,
$\beta^\alpha \to \beta^\alpha + \eta^\alpha$. In
particular,\footnote{We thank Michael Haack and Marco Zagermann
for helpful conversations on this point.} if ${\cal F}$ is turned
on in the world volume of a brane wrapping a 4-cycle $\Sigma^d$,
and if its expansion in terms of basis harmonic 2-forms is ${\cal
F} = f_\alpha \omega^\alpha$, then $\eta^\alpha = k^{\alpha \beta
d} f_\beta$, where $k^{\alpha\beta\gamma}$ denotes the
intersection number for a triplet of 2-cycles \cite{HKLVZ}.
\end{itemize}

The significance of the gauged shift symmetry is that it implies
that the positive magnetic energy (which is proportional to the
integral of ${\cal F}_{mn} {\cal F}^{mn}$ over the $D7$ volume) is
captured by a supersymmetry-breaking $D$-term in the low-energy 4D
supergravity. When nonzero this energy breaks supersymmetry in the
4D theory, just as does the magnetic flux in the underlying brane
picture. The situation becomes more complicated should other
multiplets, $Q^x$, also exist that are charged under this
symmetry. Such scalars complicate the picture because they must
also appear in the corresponding $D$ term potential, and typically
prefer to adjust their expectation values to try to cancel out the
magnetic energy and thereby restore the supersymmetry broken by
the flux. Furthermore, such scalars are often required to exist,
either by anomaly-cancellation arguments or by gauge invariance if
the axion fields should appear in the low-energy superpotential
\cite{bkq,HKLVZ,bkqplus}.

%%%%%%%%%%%%%%%%%%%%%%%%%%%%%%%%%%%%%%%%%%%%%%%%%%%%%%%%%%%%%%%%%%%%%
\subsection{The low-energy supergravity}

The interactions of these complex moduli with one another and with
gravity are described at low energies by an effective 4D theory,
that is close to an $\mathcal{N}=1$ supergravity provided that the
supersymmetry-breaking effects of the compactification to 4D are
sufficiently weak. As such it is characterized by specifying its
K\"ahler potential, $K$, its holomorphic superpotential, $W$, and
gauge kinetic function, $f_{ab}$.

\subsubsection*{K\"ahler potential}

The K\"ahler potential for the leading order 4D supergravity has
the general Type IIB form,
\be \label{KasfnofV}
    K = -2 \ln \V \,,
\ee
where $\V$ is the Calabi-Yau volume in units of the string length,
$l_s = 2\pi \sqrt{\alpha'}$. When expressed in terms of the
decompositions, $t_\alpha$, of the K\"ahler form in terms of a
basis of 2-cycles, $J = t_\alpha \omega^\alpha$, the volume
becomes $\V = \frac16 k^{\alpha\beta\gamma} t_\alpha t_\beta
t_\gamma$, where $k^{\alpha \beta \gamma}$ denotes the appropriate
intersection number for the basis 2-cycles.

For use in the supergravity action the above expression for $\V$
must be expressed in terms of the complex coordinates, $T^\alpha =
\xi^\alpha + i \beta^\alpha$, and the complex position, $z$, of
the $D3$ brane, corresponding to the chiral scalars of the
effective theory. The expression for $\V$ in terms of $T^\alpha$
and $z$ can be obtained explicitly in the case of $K3 \times
T_2/Z_2$. Because this is a product geometry its volume
factorizes,
\be \label{Vasfnofst}
    \V = \frac12 \, k^{ij} \; t_s \, t_i  t_j \,,
\ee
and is linear in the volume, $t_s$, of the torus. Here $\{
t_\alpha \} = \{t_s, t_i \}$, and $k^{s\,ij} = k^{ij}$ is a known
matrix that describes the intersection numbers of 2-cycles within
$K3$ \cite{K3}.\footnote{We adopt the convention that Greek
indices $\alpha,\beta,\cdots$ run over moduli of $K3\times
T_2/Z_2$; mid-alphabet Latin indices, $i,j,\cdots$, label only the
moduli of $K3$ (and not $t_s$ or $\xi^s$), while early-alphabet
Latin indices, $a,b,\cdots$, collectively denote $\xi^s$ and $z$.
Capitalized indices, $A,B,\cdots$ generically denote all moduli
together.} In principle, the sum on $i,j$ is over all of the
independent 2-cycles on $K3$ and so runs from 1 to 22. However we
can imagine some of the corresponding moduli to have been
stabilized (by fluxes or nonperturbative effects) at energies that
are hierarchically large compared with those of later interest for
inflationary dynamics, and in this case $i,j$ range only over the
number of remaining moduli that are lighter than these.

When restricted to a single $D3$ moving only in the toroidal
directions, the relation between the $\xi^\alpha$ and the
$t_\alpha$ becomes \cite{jockers}
\be
    \xi^i = \frac{\partial \V}{\partial t_i} = k^{ij} \; t_s t_j
    \qquad \hbox{and} \qquad
    \xi^s = \frac{\partial \V}{\partial t_s} + \omega(z,\bar z)
    = \frac12 k^{ij} \, t_i t_j + \omega(z,\bar z)\,,
\ee
where $\omega(z,\bar z)$ is the K\"ahler form on the 2-torus
(whose explicit form is given below). Inverting these expressions
for the torus volume, $t_s$, and the $K3$ 2-cycle volumes, $t_i$,
\be
    t_s = \left( \frac{k_{ij} \xi^i \xi^j}{X} \right)^{1/2}
    \quad \hbox{and} \quad
    t_i = \frac{k_{ij} \xi^j}{t_s} \,,
\ee
with $k^{ij} t_i t_j = X = 2[\xi^s - \omega(z,\bar z)]$ and
$k^{ij} k_{jk} = {\delta^i}_k$. Using these in
eqs.~\pref{KasfnofV} and \pref{Vasfnofst}, and dropping additive
constants in $K$, gives
\be
    K = - \ln X - \ln Y
\label{K1}
\ee
where $Y = \frac12 \, k_{ij} \xi^i \xi^j$. In terms of the complex
fields, $T^i = \xi^i + i\beta^i$ and $S = T^s = \xi^s + i
\beta^s$, we have
\begin{equation}
 X = S + \overline S - 2 \omega(z,\bar z) \quad \hbox{and} \quad
 Y = \frac18 k_{ij} (T^i + \overline T^i)(T^j + \overline T^j) \,.
\end{equation}

The first K\"ahler derivatives then are $K_A = \partial_A K$ (with $A
= z,S,T^i$):
\be
    K_S = - \frac{1}{X} \,, \qquad
    K_z = \frac{2\omega_z}{X} \quad \hbox{and} \quad
    K_i = - \frac{k_{ij} \xi^j}{2Y} \,,
\ee
where $\omega_z = \partial_z \omega$. The K\"ahler metric becomes
\be
    K_{A\overline B} = \left(%
 \begin{array}{cc}
  K_{a \bar b} & 0 \\
  0 & K_{i \bar\jmath} \\
 \end{array}%
 \right)
\ee
with ($a,b = S,z$)
\be
    K_{a\bar b} = \left(%
 \begin{array}{cccc}
  {1}/{X^2} &&& - {2\,\omega_{\bar z}}/{X^2}  \\
  - {2\,\omega_z}/{X^2} &&& (2 \, \omega_{z\bar z} X + 4 \omega_z
  \omega_{\bar z})/X^2 \\
 \end{array}%
 \right)
\ee
and
\be
    K_{i \bar\jmath} =  \frac{- k_{ij} Y + k_{ik} k_{jn}
    \xi^k \xi^n}{4Y^2} \,.
\ee

These have inverses
\be
    K^{\overline B A} = \left(%
 \begin{array}{cc}
  K^{\bar b a} & 0 \\
  0 & K^{\bar\jmath i} \\
 \end{array}%
 \right)
\ee
with
\be
    K^{\bar b a} = \left(%
 \begin{array}{cccc}
  X(2\,\omega^{z\bar z} \omega_z \omega_{\bar z} + X) &&&
  X \omega^{z\bar z} \omega_{\bar z} \\
  X \omega^{z \bar z} \omega_z &&&
  \frac12 X \omega^{z \bar z} \\
 \end{array}%
 \right)
\ee
and
\be
    K^{\bar\jmath i} =  4 (- k^{ij} Y +
    \xi^i \xi^j ) \,.
\ee

In terms of the real and imaginary parts of the torus coordinate,
$z = z_1 + i z_2$, and complex structure modulus, $\tau = \tau_1 +
i\tau_2$, the K\"ahler form on the torus is
\be \label{omegaT2}
    \omega = -\frac{ic (z - \bar z)^2}{2(\tau - \bar\tau)}
    = -\frac{c(z-\bar z)^2}{4\tau_2}
    = \frac{c z_2^2}{\tau_2} \,,
\ee
where $c$ is a constant to be determined below. Its derivatives
become $\omega_z = - \omega_{\bar z} = -ic(z - \bar z)/(\tau
-\bar\tau)$, $\omega_{z \bar z} = ic/(\tau - \bar\tau)$ and so
$\omega^{z \bar z} = (\tau-\bar\tau)/(ic)$, $\omega^{z \bar z}
\omega_{\bar z} = z-\bar z$ and $\omega^{z \bar z} \omega_z
\omega_{\bar z} = -ic(z-\bar z)^2/(\tau-\bar\tau) = 2\,\omega$.

Finally, there are two further properties of $K$ worth special
mention. First, as is shown in Appendix \ref{App:Noscale}, this
K\"ahler potential satisfies the no-scale identity
\be \label{noscale}
    K^{\alpha \overline \beta} K_\alpha
    K_{\overline \beta} = 3 \,.
\ee
Second, $K$ displays the periodicity of the underlying torus,
although in a subtle way \cite{uplifting}. In particular, since $X
= 2[\xi^s - \omega] = S + \overline S - 2\omega$ is explicitly
periodic under the shifts $z \to z+1$ of the torus, $K$ also
shares this property. Similarly, eq.\ \pref{omegaT2}, shows that
$K$ and $X$ are also invariant under $z \to z + \tau$ provided at
the same time we shift
\be \label{SKshift}
    S \to S - ic(2z+\tau) \,.
\ee

\subsubsection*{Holomorphic Functions}

Full specification of the low-energy 4D supergravity also requires
its holomorphic superpotential, $W$, and gauge kinetic functions,
$f_{ab}$.

\medskip\noindent{\it Gauge kinetic function}

\medskip\noindent
The gauge kinetic function, $f_{ab}(S,z)$, may be computed as a
threshold effect when computing open-string loops \cite{bhk}, or
as the classical back-reaction of the $D3$ on the $D7$ geometry in
the dual closed string picture \cite{baumann}. For the $D7$'s
located at fixed point $r$ in $K3\times T_2/Z_2$ either approach
gives $f_{ab,r} = f_r \delta_{ab}$, where (up to $z$- and
$S$-independent quantities)
\be \label{frexpr}
    f_r = S - \frac{1}{a} \Bigl\{  \ln \vartheta_1[ \pi (z_r -
    z)|\tau] + \ln \vartheta_1 [\pi (z_r + z)|\tau] \Bigr\}
    + \mathfrak{f}_r\,.
\ee
where the four fixed points on $T_2/Z_2$ are located at $z_r = 0,
\frac12, \frac12 \tau$ and $\frac12(1+\tau)$, and $\mathfrak{f}_r$
denotes a potential contribution that is independent of the fields
$S$, $T^i$ and $z$ \cite{HKLVZ}. Typically $a = 2\pi$, and
$\vartheta_1$ denotes a Jacobi theta function, for which our
conventions are specified in Appendix \ref{s:periodicity}.

\medskip\noindent{\it Superpotential}

\medskip\noindent
The appearance of a superpotential is the hallmark of an
underlying stabilization mechanism, and in the present instance we
envision the stabilization to give
\be \label{Wform}
    W = W_0 + w(S,z) + \sum_i B_i \, e^{-b_i T^i} \,.
\ee
Here $W_0$ is contributed by the flux compactification and so is
completely independent of the K\"ahler moduli. $|W_0|$ must be
chosen as small as the various nonperturbative terms in $W$ in
order to trust the shape of the potential while neglecting
corrections to $K$.

The $T^i$-dependent terms are imagined to be generated by
euclidean $D3$ branes ($ED3\,$s) wrapped about the torus together
with one of the various 2-cycles of $K3$ \cite{ak}, in which case
$b_i = 2\pi$. The coefficients $B_i$ could depend on the position
of the $D3$ in the $K3$ directions, and we imagine this dependence
to have provided the forces which prevent $D3$ motion in these
directions (allowing the neglect of this motion in what follows).

Wrapping such $ED3$ branes about the $K3$ similarly can stabilize
its volume, as can gaugino condensation on the $D7$'s localized at
the fixed points of the torus. (Which of these obtains depends on
the details of the fluxes that are applied \cite{GKTT}.) This is
what generates the $S$-dependent term, $w(S,z)$, of
eq.~\pref{Wform}, which is predicted to take the following
form\footnote{Notice that our definition of $a$ does not contain
the factor of $1/N_r$ in the case of gaugino condensation on the
$D7$ stack at fixed point $r$; our notation differs from that of
ref.\ \cite{haack} in this respect.\label{anote}}
\be
    w(S,z) = \sum_r w_r(S,z)
    =\sum_r \Bigl\{A_r  e^{-a f_r(S,z)} \Bigr\}^{1/N_r}
    = \sum_r \Bigl\{ e^{-a S} F_r(z,\tau) \Bigr\}^{1/N_r}
\ee
where for future notational convenience we introduce the function
\be
    F_r(z,\tau) \equiv  A_r \, \vartheta_1[\pi(z_r - z)|\tau]
    \vartheta_1[\pi(z_r +z)|\tau] \,.
\label{Freq}
\ee
Here $N_r = 1$ if $w_r(S,z)$ arises due to an $ED3$, but $N_r$
depends on the gauge group involved if $w_r(S,z)$ is generated by
gaugino condensation. For instance, $N_r = N$ if gaugino
condensation arises for an $SU(N)$ or $SO(N+2)$ gauge group.

The quantity $w$ is invariant under the shifts $z \to z + 1$ and
$z \to z + \tau$ provided that $S$ also shifts appropriately. The
transformation properties of Appendix \pref{s:periodicity} show in
particular that invariance under the transformation $z \to z+\tau$
requires $S \to S - 2\pi i(2z+\tau)/a$. Comparing this with
condition \pref{SKshift}, required for invariance of $K$, shows
that invariance of the complete scalar potential requires the
constants $c$ and $a$ must be related by \cite{uplifting}
\begin{equation} \label{acrelation}
 c = \frac{2\pi}{a} \,,
\end{equation}
and so $c=1$ if $a = 2\pi$.

Whether gaugino condensation occurs on the stack of branes at a given
fixed point depends on the low energy gauge group and field
content. We assume there is enough freedom to turn on condensation at
one or more fixed points.

%%%%%%%%%%%%%%%%%%%%%%%%%%%%%%%%%%%%%%%%%%%%%%%%%%%%%%%%%%%%%%%%%%%%
\subsection{Low-energy scalar interactions}

The low-energy scalar interactions are generically governed by
${\cal L} = {\cal L}_{\scriptscriptstyle SG} + \delta{\cal
L}_{sb}$, where ${\cal L}_{\scriptscriptstyle SG}$ denotes the
relevant part of the 4D supergravity lagrangian,
\be
    {\cal L}_{\scriptscriptstyle SG} = - \sqrt{-g} \, \Bigl[
    V_F(T,\overline T) + V_D(T, \overline T) +
    K_{A \overline B} \, \partial_\mu
    T^A \partial^\mu \overline T^{B} + \cdots \Bigr] \,,
\ee
and the SUSY-breaking term
\be
    \delta {\cal L}_{sb} = - \sqrt{-g} \, \Bigl[
    V_{\rm up}(T,\overline T) + \cdots\Bigr]
\ee
denotes the derivative expansion of any terms which cannot be put
into the 4D $\mathcal{N}=1$ supergravity form, such as low-energy
terms due to the presence of a supersymmetry-breaking anti-$D3$
brane. Any such terms must be perturbatively small in order for
the 4D supergravity form to be a good approximation, such as might
occur if the $\overline {D3}$ were localized in a strongly warped
throat.

The $F$- and $D$-term potentials are given as usual by
\be \label{VF}
    V_F = e^K \Bigl[K^{A \overline B} D_A W \overline{D_B W}
    - 3 |W|^2 \Bigr] \,,
\ee
and
\be \label{VD}
    V_D = \frac12 \sum_r \F_r^{ab} D_{a,r} D_{b,r} \,,
\ee
where the sum is over the 4 fixed points of $T_2/Z_2$ and
$\F^{ab}_r$ denotes the inverse matrix for $\hbox{Re}\, f_{ab,r}$.
The auxiliary fields, $D_{a,r}$, are given by
\begin{equation}
 D_{a,r} = \delta_a K = \frac{ \partial K}{\partial T^\alpha} \,
 \eta_{a}^\alpha
 + \frac{\partial K}{\partial Q^x} \, (t_a Q)^x
 + (\overline Q t_a)_x \frac{\partial K}{\partial
 \overline Q_x} \,  \,,
\end{equation}
where $t_a$ denotes the appropriate gauge generator acting on any
low-energy charged chiral fields, $Q^x$, that happen to be present
(often arising as low-energy open string states). The quantity
$\eta^\alpha$ denotes the shift of the moduli fields, whose
imaginary parts transform as $\delta_a \beta^\alpha =
\eta_{a}^\alpha$. Such shifts arise when the corresponding axionic
shift symmetry is gauged by background 2-form fluxes localized on
$D7$ brane, as described in more detail in section
\ref{sec:fieldcontent} above. Notice in particular that for fluxes
localized on the $D7$'s wrapping the $K3$, $\eta^\alpha$ never
points in the direction of the $K3$ volume modulus, $S$: $\eta^s =
0$. This follows from the vanishing for $K3 \times T_2/Z_2$ of all
intersection numbers of the form $k^{\alpha s s} = 0$.

The $z$-dependence of the $D$-term potential, eq.\ \pref{VD}, has
a simple physical implication. If $D_{a,r} = 0$ after the $Q^x$
are minimized, then $V_D$ is $z$-independent. If $D_{a,r} \ne 0$,
on the other hand, the $z$-dependence of $V_D$ arises from the
gauge coupling function, $f_r$. For small $D3$-$D7$ separation this
varies logarithmically, with $\hbox{Re} \, (f_r - S) \propto -
\ln|z-z_r|$. This describes a force acting on the $D3$ due to the
$D7$'s that vanishes (by supersymmetry) in the absence of the
magnetic flux, but is otherwise nonzero. Furthermore, this force
arises due to tree-level closed-string exchange (since the
$z$-dependence of $f_r$ arises due to the classical back-reaction
on the bulk geometry by the $D3$ brane \cite{baumann}).
Equivalently, because of open-closed string duality, this force
can be regarded as being due to open string loops, and as such can
be regarded as the 4D supergravity description of the
`Coleman-Weinberg' part of the $D3$-brane potential used for
inflationary purposes in ref.~\cite{haack}.

The difficulty with using $V_D$ is that the energetics of the
charged fields, $Q^x$, if present, usually prefers them to adjust
to ensure $D_r = 0$ ref.\ \cite{bkq,HKLVZ,bkqplus}, thereby
turning off the flux-induced $D3$-$D7$ force. This is the
low-energy 4D supergravity's way of expressing how the $D7$'s can
prefer to adjust internally to preserve supersymmetry, and thereby
eliminate the $D3$-$D7$ Coleman-Weinberg interaction. Furthermore,
such charged field very often {\it must} exist. They are typically
required, for instance, to understand the gauge invariance of $W$
once $W$ depends --- such as in eq.~\pref{Wform} --- on fields
like $T^i$ if these shift under a gauge symmetry.

\subsubsection*{SUSY-breaking terms}

Following KKLT \cite{KKLT} we take the contribution of any
anti-$D3$ branes (should these be present) to be perturbatively
small and contained in $V_{\rm up}$, whose detailed form depends
on whether or not the antibrane is located in a strongly warped
region. Warped type IIB compactifications of $K3 \times T_2/Z_2$
were examined in ref.\ \cite{dasgupta}.

The 4D potential due to the tension of an anti-$D3$ brane is (in
the 4D Einstein frame)
\begin{equation}
 V_{\rm up} = \frac{\hat E \, e^{4 \A}}{\V^2} \,,
\end{equation}
where the constant $\hat E$ is proportional to the $D3$ tension,
$T_3$. The warp factor, $\A$, is defined by the form of the string
frame metric,
\be
 \d s^2_{10} = e^{2\A} \eta_{\mu\nu} \d x^\mu \d x^\nu +
 e^{-2\A} g_{mn} \d y^m \d y^n \,,
\ee
where $g_{mn}$ is the metric of $K3 \times T/Z_2$ such that $\V =
\int \sqrt{g_6}\, \d^6 y$. To leading order the warp factor $\A$
depends only on the $K3$ coordinates, although this changes once
one includes corrections in $\alpha'$ and the string coupling,
$g_s$ \cite{dasgupta}.

When evaluated in a strongly warped throat it happens that
$e^{4\A} \propto e^{-\zeta} \V^{2/3}$, where $\zeta = 8\pi
n_1/(3g_s n_2)$ is a combination of certain integer flux quantum
numbers, $n_i$, and so the total volume-dependence of an antibrane
uplifting potential is
\be \label{uplift}
    V_{\rm up} = \frac{E}{\V^p} \,,
\ee
with $E = \hat E e^{-\zeta}$ and $p = 4/3$ (or $E = \hat E$ and $p
= 2$) if the anti-$D3$ is (is not) located in a warped throat. In
the absence of a better understanding of the cosmological constant
problem we imagine $E$ to be tuned to ensure that the scalar
potential vanishes at its minimum.

Locating the antibrane within a warped throat has several
well-known advantages.
\begin{itemize}
\item Warping suppresses the scale of the supersymmetry-breaking
physics relative to other scales, and this helps to sequester its
effects from the SUSY-breaking sector \cite{warpSUSY}. This is
required to justify regarding the SUSY-breaking terms of $\delta
{\cal L}_{sb}$ as small perturbations to the 4D supergravity
action.
\item Warping allows the scale of the uplifting to be tuned in
small steps, potentially allowing a closer approach to a vanishing
potential at the minimum.
\item Warping decreases the Coulomb potential between the $D3$ and
$\overline {D3}$, largely because it suppresses the
$\overline{D3}$ charge. This is important for inflationary
applications because without the warp factor the Coulomb force
tends to ruin slow roll for $D3$ motion in the $z$ directions.
Asymmetric compactifications with the $K3$ radius much larger than
the torus radius do not improve this situation \cite{BBar}.
\item Finally, warping tends to localize the $\overline {D3}$ by
making it settle into the bottom of the throat. This keeps it from
migrating to one of the branes and perhaps annihilating.
Furthermore, although the $D3$ {\it is} mobile, we imagine that
the stabilization of the $K3$ moduli in $V_F$ fixes its position
within $K3$ (see the discussion below eq.~\pref{Wform}), and does
so far from the throat.\footnote{An explicit construction with the
$D3$ stabilized away from the throat's tip can be found in ref.\
\cite{baumann2}.} This keeps the $D3$ from migrating to the
$\overline{D3}$ and annihilating, leaving it free to play an
inflationary role as it moves along the torus.
\end{itemize}

%%%%%%%%%%%%%%%%%%%%%%%%%%%%%%%%%%%%%%%%%%%%%%%%%%%%%%%%%%%%%%%%%%%%
\section{The inflationary model}

To make the search for inflationary solutions manageable we
imagine all but one of the moduli $T^i$ to be stabilized with
masses larger than those relevant for the inflationary motion,
allowing us to specialize the previous setup to only three complex
fields: the $K3$ volume, $S$, the $D3$ position on the torus, $z$,
plus the one remaining modulus $T$. Our motivation for keeping one
of the $T$'s is to maintain the no-scale form of the low-energy
supergravity, whose K\"ahler potential (up to an irrelevant
additive constant) then is
\be
    K = - \ln\Bigl[S + \overline S - 2\,\omega(z,\bar z) \Bigr]
    - 2 \ln(T + \overline T) \,.
\ee
Although we follow $T$ numerically when searching for inflationary
dynamics, it turns out to play a negligible role in the actual
inflationary slow rolls we eventually find.

To simplify the notation in this section we denote the real and
imaginary components of the fields $S$ and $T$ by $\xi^s = s$ and
$\xi^t = t$, so
\begin{equation}
 S = s + i \alpha, \quad T= t + i \beta
\end{equation}
while as before, $z= z_1+ i z_2$.

The superpotential, eq.~\pref{Wform}, for this reduced theory
becomes
\be
    W = W_0 + \sum_r w_r(S,z) + B e^{-b T} \,,
\ee
where $w_r = \left[ F_r(z)e^{-aS} \right]^{1/N_r}$. For simplicity
we restrict in what follows to the case where $N_r = N$ is
independent of $r$. Then the $S$-dependent part of $W$ becomes
\be
    w(S,z) =  A(z) \, e^{-a S/N} \label{W1easy}
\ee
with $A(z) = \sum_r F_r(z)^{1/N}$. Notice that the scalar
potential derived from this superpotential is periodic under
$\alpha \to \alpha+2\pi N/a$ and $\beta \to \beta+2\pi/b$.

%%%%%%%%%%%%%%%%%%%%%%%%%%%%%%%%%%%%%%%%%%%%%%%%%%%%%%%%%%%%%%%%%%%%
\subsection{F-term potential}

The $F$-term potential is found by specializing the earlier
results to the three fields of interest. The derivatives of the
superpotential are
\be
 W_z = \sum_r \frac{w_r}{N_r}  (\partial_z \ln F_r)
 = w \, \partial_z \ln A \,, \qquad
 W_S = -a \sum_r \frac{w_r}{N_r} = - \frac{a \,w}{N} \,, \qquad W_T = -b B
 e^{-bT} \,,\label{dW}
\ee
and (keeping in mind the no-scale form of the K\"ahler potential)
the $F$-term potential becomes
\ba
    V_F &=& \frac1{X (T+\bar T)^2} \left\{ \sum_{A=S,T,z}
    K^{A\bar A} \Bigl[ W_A \overline{W}_A + (K_A W
    \overline{W}_A +c.c.) \Bigr] \right. \nn\\
    && \qquad \qquad \left. \phantom{\sum_S} +
    \Bigl[ K^{S\bar z} (W_S \overline{W}_z + K_S W \overline{W}_z
    + K_{\bar z} W_S \overline W +c.c) \Bigr] \right\} \,,\label{V1}
\ea
where $X = 2[s - \omega(z,\bar z)]$. Notice that for large $K3$
volume, $s$, we have $X \approx 2 s$ and so when all else is equal
it is the $K^{S\overline S} \propto s^2$ term that dominates.

%%%%%%%%%%%%%%%%%%%%%%%%%%%%%%%%%%%%%%%%%%%%%%%%%%%%%%%%%%%%%%%%%%%%%%
\subsubsection*{Axion minimization}

The axion fields, $\alpha$ and $\beta$, can now be minimized
explicitly. The only terms involving these fields come from $V_F$
and are given by
\begin{equation}
    V_{ax} = \frac1{X (T+\bar T)^2} \left\{ K^{T\overline T}
    K_T \overline{W}_T (W_0 + w) + K^{b \bar a} K_b \overline{W}_a
    \left( W_0 + B e^{-bT} \right)
    +c.c. \right\} \,,\label{Vax} \nn
\end{equation}
where, as before, the indices $a,b = S,z$. This contains terms
proportional to $\cos(b\beta)$, $\cos(a \alpha/N)$ and
$\cos(b\beta - a\alpha/N)$. It is convenient to use an overall
phase rotation to choose $W_0$ to be real and negative, since for
the parameter range of later interest this ensures that ${\rm
Im}\,T = \beta = 0$ at its minimum. Minimizing $\alpha$ similarly
amounts to replacing $A \to |A|$ in the remaining equations.

%%%%%%%%%%%%%%%%%%%%%%%%%%%%%%%%%%%%%%%%%%%%%%%%%%%%%%%%%%%%%%%%%%%%%%%
\subsubsection*{The supersymmetric AdS minimum}

$V_F$ as described above has a supersymmetric AdS minimum,
corresponding to the solutions to $D_A W = 0$. The condition $D_S
W = D_T W = 0$ to be solved for $s = s_0(z)$ and $t = t_0(z)$ may
be written as
\begin{eqnarray}
\label{Beq}
 |B| &=& \frac{a|A(z)| X_0}{N b \, t_0} \, e^{bt_0 - as_0/N}\,, \\
 W_0 &=& - |A(z)|e^{-a s_0/N} \[ 1+ \frac{aX_0}{N}
 \left(1+\frac{1}{b\,t_0} \right) \]  \,, \label{susysol}
\end{eqnarray}
where $X_0 = 2[s_0 - \omega(z,\bar z)]$. Using these conditions in
$D_z W = 0$ then implies $z$ must satisfy
\begin{equation}
 |A(z)| e^{-a s_0/N} \left[ \partial_z \ln |A(z)|
 + \frac{2\pi(z-\bar z)}{N\tau_2} \right] =0 \,,\label{DzW}
\end{equation}
which uses the explicit form, eq.~\pref{omegaT2}, of $\omega$ as
well as the condition $ac = 2\pi$, eq.~\pref{acrelation}.

Eq.~\pref{DzW} is always solved by $A(z) = 0$, but in this case
$w(S,z) = 0$ and so the $S$-modulus is not stabilized. If
$S$-stabilization occurs at a single fixed point, $r_0$, then
$A(z) = [F_{r_0}(z)]^{1/N}$ can vanish when the $D3$ approaches $z
= z_r$. Similar solutions also exist for $ED3$'s located at all
four fixed points, for which $A = \sum_r F_r$, since in this case
$A=0$ when $(z_1,z_2) = (1/4+n/2,0)$ for $n$ an arbitrary integer.
(These last solutions are most easily seen in the limit $\tau_2
\gg 1$, for which Appendix \ref{s:theta} shows $\sum_r F_r \propto
\cos(2\pi z)$ -- see eq.\ \eref{largetau}.)

To obtain a supersymmetric extremum without destabilizing $S$
requires the bracket in \eref{DzW} to be zero. When $A =
F_{r_0}^{1/N}$ these extrema are at $z = (n/2, m\tau_2/2)$ (which
includes as special cases the points where $A(z) = 0$). When $A =
\sum_r F_r$ the solutions instead are $(z_1,z_2) = (n/2,0)$, and
$(n/4,\tau_2/2)$, with $n$ an integer. (Again these latter
solutions are simplest to see in the large $\tau_2$-limit.) Which
of these are maxima or minima depends on the parameters used (and
in any case can change after including an uplift term, as we shall
see).

The potential at this minimum becomes
\be
 V_F^{\rm AdS} = - \frac{3 |W_{min}|^2}{X (T+\overline T)^2}
 = - \frac{3\,a^2|A|^2X_0}{4N^2 t_0^2} \, e^{-2a s_0/N}
 \,. \label{AdSmin}
\ee
For $\tau_2 \gg 1$, evaluation of the potential near this minimum
shows it to be flattest along the $z_2$ direction, while for
$\tau_2 \ll 1$ it is instead flatter along $z_1$.

%%%%%%%%%%%%%%%%%%%%%%%%%%%%%%%%%%%%%%%%%%%%%%%%%%%%%%%%%%%%%%%%%%%%%%%
\subsection{Uplifting}

We next consider lifting this solution to positive values of the
potential at the minimum, using either a $D$-term potential or
that of an anti-$D3$ brane.

\subsubsection*{D-term potential}

If a $D$-term potential due to magnetic flux located at a brane at
fixed point $z_r$ were to exist, either due to the absence of
charged matter fields, $Q^x$, or if their complete potential is
minimized at $Q^x = 0$, it would depend on $X$ and $T$ in the
following way: $V_{D,r} \propto (K_\alpha \eta^\alpha)^2/{\rm
Re}(f_r)$. Here $\eta^\alpha$ measures the linear combination of
axion fields which is gauged by the magnetic flux in question,
which the discussion of previous sections shows never points in
the $S$ direction. So for the fields of present interest $K_\alpha
\eta^\alpha \propto K_T \propto 1/t$, and so
\be \label{Dterm3fd}
 V_D^r(S,z) = \frac{E_r}{{\rm Re}(f_r) t^2} \,,
\ee
where $E_r$ is a constant.

But there is a consistency problem with having $T$ shifting in
this way under a $U(1)$ symmetry without also having charged
fields $Q^x$ be present. The problem is that if $T$ shifts in the
way required to appear in $D$, then this same symmetry precludes
the existence of a term in the superpotential like $B e^{-bT}$, as
was required to stabilize $T$. Charged fields like $Q^x$ can
resolve this kind of paradox because their presence in $W$ can
combine with $T$ to make the superpotential invariant. We refer
the reader to refs.~\cite{HKLVZ,bkqplus} for more detailed
discussions of these issues.

Because of this issue, we perform our main search for inflation
using an alternative source of uplifting, such as from an
anti-$D3$ brane. This is what we use in our most successful
inflationary scenario, described below. However, following
\cite{haack} we also seek inflation using eq.~\pref{Dterm3fd}, in
the spirit that it might ultimately turn out to capture the
low-energy dynamics of some better motivated, but more
sophisticated, string constructions. In particular, we use this
form of uplifting when exploring the limit where the $D3$ and
$D7$'s are in close proximity, in order to try to follow as closely
as possible the analysis of ref.~\cite{haack}.

%%%%%%%%%%%%%%%%%%%%%%%%%%%%%%%%%%%%%%%%%%%%%%%%%%%%%%%%%%%%%%%%%%%%%%%%%
\subsubsection*{Anti-$D3$ brane}

When uplifting with an anti-$D3$ brane, we assume the potential to
depend only on the volume, with the form discussed above
\begin{equation}
 V_{\overline {D3}} = \frac{E }{\V^p} = \frac{E}{X^{p/2}
 (T+\overline T)^p} \label{generalp}
\end{equation}
which uses $\V = \sqrt{X} (T+\bar T)$. As before, the power is
$p=2$ for anti-branes in unwarped regions and $p=4/3$ when the
antibrane is deep within a warped throat \cite{KKLMMT}, and so we
use $p=4/3$ in our main search for inflationary solutions. We have
checked that similar solutions also exist when $p=2$, however.

Since neither $X$ nor ${\rm Re}(f_r)$ depend on the axions,
$\alpha,\beta$, an uplifting potential of either $D$-term or
antibrane type would not alter their minimization.

%%%%%%%%%%%%%%%%%%%%%%%%%%%%%%%%%%%%%%%%%%%%%%%%%%%%%%%%%%%%%%%%%%%%%%%%%%%%
\section{Slow-Roll Inflation}

We next search the potential for the fields $S$, $T$ and $z$,
seeking slow-roll regions for which the effective single-field
slow-roll parameters, $\epsilon$ and $\eta$, can be made small. We
find that inflation does not generically arise, but -- as for many
other brane-inflation models -- slow roll can occur provided some
of the parameters in the potential are mildly tuned (see, however,
\cite{KMI} for potentially less tuned alternatives). In this
section we describe two such examples.

We first search for slow-roll regimes that do not rely on the
existence of a $D$-term potential, by uplifting using an
antibrane. We find that slow-roll inflation is possible to obtain
near a saddle point, where we use a superpotential generated by
gaugino condensation localized at a single fixed point on the
torus, with the $D3$ located as far away as possible from this
fixed point. In order to achieve inflation the shape of the torus
must be tuned to be very close to square, to within a part in
$10^4$, but once this is done the resulting slow roll is largely
insensitive to the other parameters in the $F$-term potential.

The second example we present is a direct analog of the scenario
proposed in ref.~\cite{haack}, with the $D3$ very close to one of the
$D7$ fixed points. Two $D$-terms are added in this setup, one to drive
inflation and the other to uplift the potential. In doing so we follow
\cite{haack} and put aside the concerns given above whether the
charged matter fields cause the $D$-term to relax to zero. We perform
our search numerically, using the full expressions for $V_D$ and
$V_F$, rather than searching analytically using a simplified
parametrization of the potential.  Taking the inflaton to be primarily
in the $z_1$-direction, we do not find any example that resembles
standard $D$-term driven hybrid inflation (and we identify the reasons
for this difference with \cite{haack}). Instead, we find that
inflation can occur at an inflection point of the potential, for which
the $D$-term and $F$-term contributions to $\epsilon =0$ and $\eta$
are fine-tuned to be small.

\FIGURE[ht]{ \epsfig{file=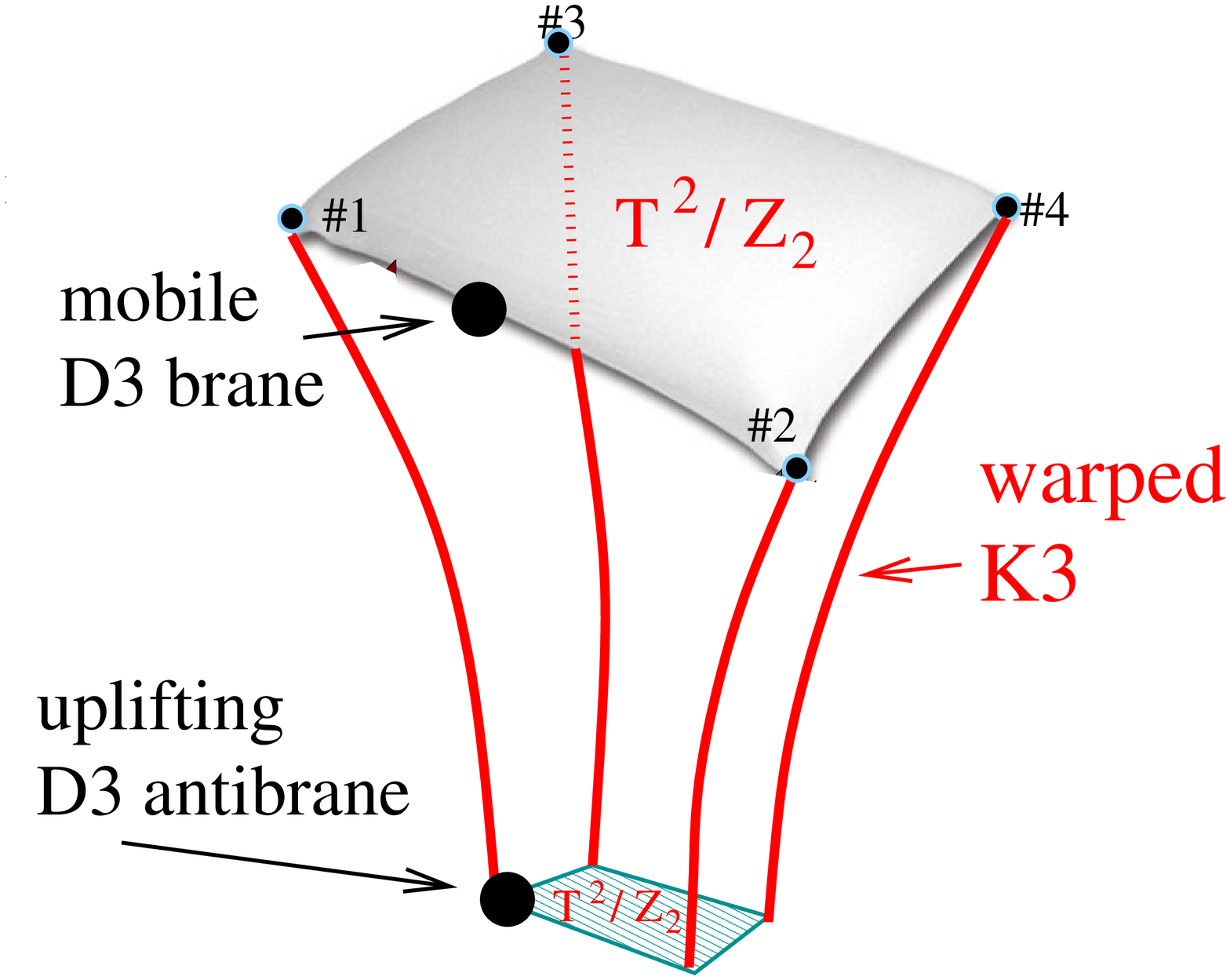,width=0.5\hsize}
\caption{Illustration of warped $\overline{D3}$ uplifting with
$D3$ confined to top of throat.} \label{warped-d3d7} }

%%%%%%%%%%%%%%%%%%%%%%%%%%%%%%%%%%%%%%%%%%%%%%%%%%%%%%%%%%%%%%%%%%%
\subsection{Axionic inflation}

We start with our best example of $D3$-$D7$ inflation. We imagine
gaugino condensation to occur only at a single fixed point, $z_r
=0$, leading to a superpotential term as given by \eref{W1easy},
\begin{equation}
 W = W_0 + \left[ A_0 e^{-aS} \vartheta_1^2(\pi z|\tau)
 \right]^{1/N} + B e^{-bT}
\end{equation}
which absorbs a conventional sign into the constant $A_0$.

We take the uplifting potential provided by a $\overline{D3}$,
localized in a throat, leading to a contribution as in
eq.~\pref{generalp} with $p=4/3$. (Although we find the warped
version of this picture most appealing, we have also checked that
inflation can work without warping.) Finally, as discussed at
length above, we imagine the $D3$ cannot move in the $K3$
directions but is mobile within the $T_2$. We assume the $D3$ is
not near the throat, so the uplifting potential is the only
antibrane perturbation to the $D3$ motion. Notice that this
uplifting depends on the $D3$ position through the factor
$X=2[{\rm Re} S - \omega(z, \bar z)]$ of eq.~\eref{generalp}, and
this plays an important role in the shape of the $D3$ potential.
Although we believe this construction --- illustrated in figure
\ref{warped-d3d7} --- to be plausible, we leave a detailed
derivation for future work.

For the warped configuration, it is straightforward to find an
almost-flat saddle point in the potential, for any values of the
superpotential parameters $a,b,A_0,B$. This is done simply by
tuning the single parameter $\tau_2\equiv$ Im($\tau$), which
determines the shape of the torus. Setting $\tau_1\equiv$
Re($\tau)=0$ for simplicity, we find that if $\tau_2$ is close to
1 (so that $T_2/Z_2$ is nearly square) we get a flat potential
close to the antipodal point, $z = \frac12(1 + \tau)$, of the
fixed point source at $z=0$. The surprise is that the unstable
direction turns out not to be in the $z_1$-$z_2$ plane, but rather
is a linear combination of $z_1$ and $\alpha = \hbox{Im}\, S$, the
axion associated with the volume modulus.

An explicit example leading to an inflationary slow roll is given by the
parameter choices
\begin{equation}
% W_0 = -10^{-4},\  a=1,\ A=1,\ B=1,\ b=1,\ N = 4,\ \tau_1=0
 W_0 = -4.14\times 10^{-7},\  a=b=2\pi,\ A_0 =0.538,\ B= 0.912,\
 N = 4,\ \tau_1=0,\
% \quad \hbox{and} \quad
z_2 = \frac{\tau_2}{2} \,,
 \label{params}
\end{equation}
which corresponds to a minimum at $s_0 = 11.54$, $t_0 = 2.802$ and
$\beta = 0$. Uplifting requires taking $E = 1.70217\times
10^{-13}$ (when $p=4/3$). These values were chosen to satisfy the
COBE normalization of the power spectrum, $P=4\times 10^{-10}$, at
the scale which we take to be 55 $e$-foldings before the end of
inflation. For this purpose we approximate $P$ as $H^4/(50\pi^2
L_{\rm kin})$ where $L_{\rm kin}$ is the kinetic energy of the
fields.

\FIGURE[ht]{
\epsfig{file=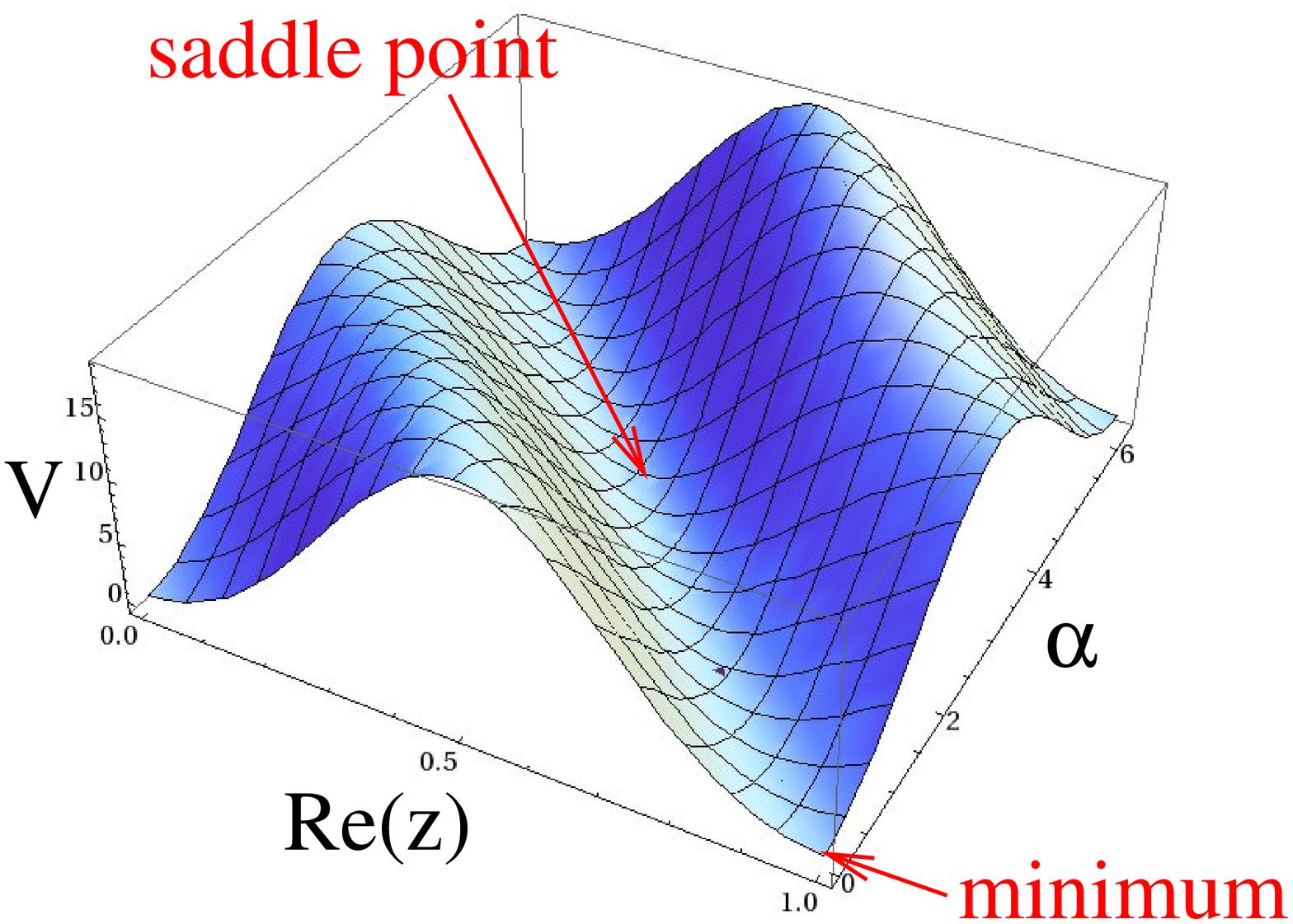,width=0.5\hsize}\epsfig{file=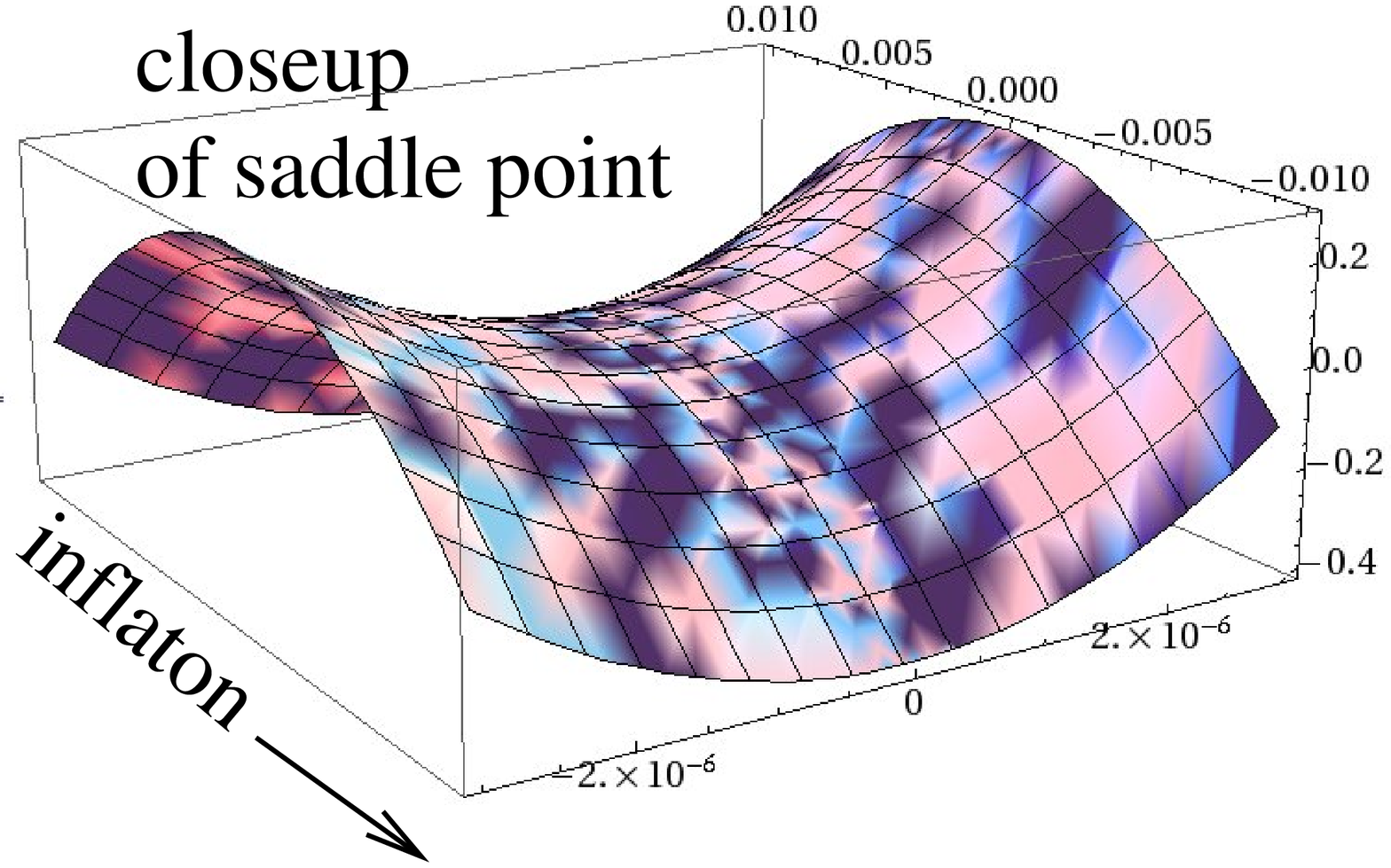,width=0.5\hsize}
\caption{Left: the potential for the warped uplifting saddle point
in the the $z_1$-$2\pi\alpha$ plane.  Right: closeup of the saddle
point region at $z_1=1/2$, $2\pi\alpha=\pi$.} \label{Vxa} }

The resulting potential is displayed in figure \ref{Vxa} as a
function of the remaining fields $z_1$ and $\alpha$. The
inflationary saddle point at $z_1 = \alpha = \frac12$ is visible
in the figure, at which the unstable direction numerically
evaluates to
\begin{equation}
 \hat\phi = {1\over\sqrt{2}}\, \hat\alpha -{1\over\sqrt{2}}\, \hat z_1 \,,
 \label{direction}
\end{equation}
where $\hat z_1$ and $\hat \alpha$ denote unit vectors in these
two coordinate directions of field space. (The components in the
directions of the other, heavy fields are smaller than this by a
factor of $10^{-8}$.) An initial condition near this saddle point
initially moves in the direction given by eq.\ (\ref{direction}),
which is also the direction towards the local minimum at $z_1 =
\alpha = 0$. In the resulting motion the $D3$ falls from the fixed
point at $z = \frac12(1+\tau)$ to the fixed point at $z = \frac12
\tau$, driven by the nonperturbative physics situated at $z=0$.

We regard the values of $t_0$ and $s_0$ to be just within the
domain of validity of the $\alpha'$ and $g_s$ expansions, although ideally
larger values would be preferable.\footnote{It is our use of $N=4$
in $w(S,z)$ while $N=1$ for the Euclidean D3-brane superpotential
$Be^{-bT}$, together with our requirement that $A_0, B < 1$ (to
avoid having a large energy scale in the nonperturbative
superpotential) that leads to our obtaining the hierarchy
$t_0\cong s_0/4$. (Note that eq.\ (\ref{Beq}) shows that
exponentially large values of $B$ would be required to make
$t_0\sim s_0$.) On the other hand, if we take $N=1$ for $w(S,z)$
then the values $s_0\sim t_0\sim 10$ put the energy scale of the
potential far below that needed for the COBE normalization.} With
the parameters chosen we have $a s_0/N \simeq b\, t_0 \simeq 18$,
ensuring the suppression of the nonperturbative superpotential.
Furthermore $\V \simeq \sqrt{s_0} \, t_0 \simeq 10$, so known
$\alpha'$ corrections to $K$ are of order \cite{BBHL}
\begin{equation}
 \frac{\delta K}{K} \simeq \frac{\chi \zeta(3)}
 { 2(2\pi)^3 g_s^{3/2} \V \ln \V}
 \simeq \frac{0.1}{g_s^{3/2} \V \ln \V} \,,
\end{equation}
where $\chi = 48$ is the Euler number of $K3 \times T_2/Z_2$ and
$\zeta(3) \simeq 1.2$. For $\V \simeq 10$ this is of order $3/\V
\ln \V \simeq 0.1$ (or $0.6/\V \ln \V \simeq 0.03$) if $g_s \simeq
0.1$ (or $g_s \simeq 0.3$). Comparatively small values for $t_0$
and $s_0$ are driven by the requirement that the potential be
large enough to reproduce the observed primordial scalar
perturbations, and are a reflection of a common tension in brane
inflation models between this condition and the control over the
$g_s$ and $\alpha'$ expansions. We regard the present calculation
as being sufficiently accurate to demonstrate the existence of a
slow roll, motivating a more detailed search for inflation with
larger $s_0$ and $t_0$.

We remark in passing that our numerics follow all six of the
fields $s$, $t$, $\alpha$, $\beta$, $z_1$ and $z_2$, but for the
inflationary example considered here it turns out that the
variation of the heavy fields $s$, $t$, $z_2$ and $\beta$ found
numerically during inflation proved to be negligible, the largest
being 1 part in $10^6$ for $s$. This can be understood
analytically, and is consistent with the suppression of the
perturbations of these fields by their masses, which are heavy
compared with the inflaton directions. We thus find it to be a
good approximation to ignore the slight motion of the heavy fields
during inflation, even though our numerical code evolves all six
of the fields subject only to the slow-roll
approximation.\footnote{Making the slow-roll approximation in the
numerical evolution greatly reduces the computational burden,
while still giving the correct trajectory during inflation, and
also allowing sufficiently accurate determination of when
inflation ends.}

Remarkably, the potential at the saddle point is acceptably flat
for slow-roll inflation for a reasonable range of parameter values
in the vicinity of those of eq.~\pref{params}, {\it provided} we
tune $\tau_2$ to be in the range
\begin{equation}
 \tau_2 = 1.00174 - 1.00184 \,.
\end{equation}
This range corresponds to the $\eta$ parameter at the saddle point
in the interval $-0.04 < \eta_{\rm saddle} < 0$; see figure
\ref{nseta}.  The lower value of $\tau_2$ gives the larger value
of $\eta_{\rm saddle}$, and $\tau_2=1.00174$ yields 230 $e$-foldings
when starting at a displacement of $0.001$ from the saddle point.
Although $\tau_2$ must be tuned at the level of 1 part in
$10^{4}$, it is only this one parameter in the superpotential that
needs such fine adjustment. It is suggestive that the required
value for $\tau_2$ is so close to $\tau_2=1$, which corresponds to
a square torus, and although the symmetry of this geometry has
been argued to lead to special cancellations amongst inter-brane
forces \cite{BBar}, we do not have a symmetry argument for why the
inflationary value of $\tau_2$ is not precisely at 1.

If we change the parameter values in (\ref{params}), the position
of the saddle point typically shifts, as does the unstable
direction. For example with $W_0 = -10^{-6}/(2\pi)^{3/2}$,
$a=3\pi$, $A=2/(2\pi)^{3/2}$, $B=3/(2\pi)^{3/2}$, $b=\pi$ and $N =
1$ we find the saddle point moves to $z_1 = \frac12$ and $\alpha =
\frac13$, with the unstable direction becoming $\hat\phi =
0.97266\, \hat\alpha -0.23221\, \hat z_1$. However, the tuning
needed to get inflation again simply requires $\tau_2$ close to
unity; with $\tau_2 = 1.00673$ giving about $300$ $e$-foldings of
inflation.

It is also possible to get inflation from unwarped $\overline{D3}$
uplifting, where the dependence of $V_{\rm up}$ on the volume goes
like $1/\V^2$ instead of $1/\V^{4/3}$. In this case, using exactly
the same superpotential parameters as (\ref{params}), we find that
the locations of the minima and saddle points get interchanged,
with the minimum at $(z_1,\alpha) = \left( \frac12, \frac12
\right)$ and the saddle at $(z_1,\alpha)=(0,0)$ (and $z_2=\frac12
\tau_2$ as before). However the value of $\tau_2$ needed for
flatness is now farther from unity: $\tau_2 = 1.61683$. The
direction of the inflaton is exactly the same as for the
corresponding warped case, eq.\ (\ref{direction}).

Although the qualitative features of our scenario are robust to
changes in the superpotential parameters, they are on the other
hand rather sensitive to the detailed form of the uplifting
potential (\ref{generalp}). We find that the inflationary
mechanism fails if one tries to replace the $\overline{D3}$
uplifting with a D-term, for example located at the $D7$ stack at
$z=\frac12$. Either the $\eta$ parameter cannot be tuned to be
small, or the $D3$ is attracted to the $D7$ which sources the
D-term, leading to annihilation and the removal of the uplifting.
Moreover, the $z_2$-dependence appearing in the uplifting term
through $X$ is also important; neglecting this dependence leads to
an additional negative eigenvalue of the curvature matrix, along a
linear combination of the $s$ and $z_2$ directions, which spoils
the slow roll at the saddle point.

\FIGURE[ht]{ \epsfig{file=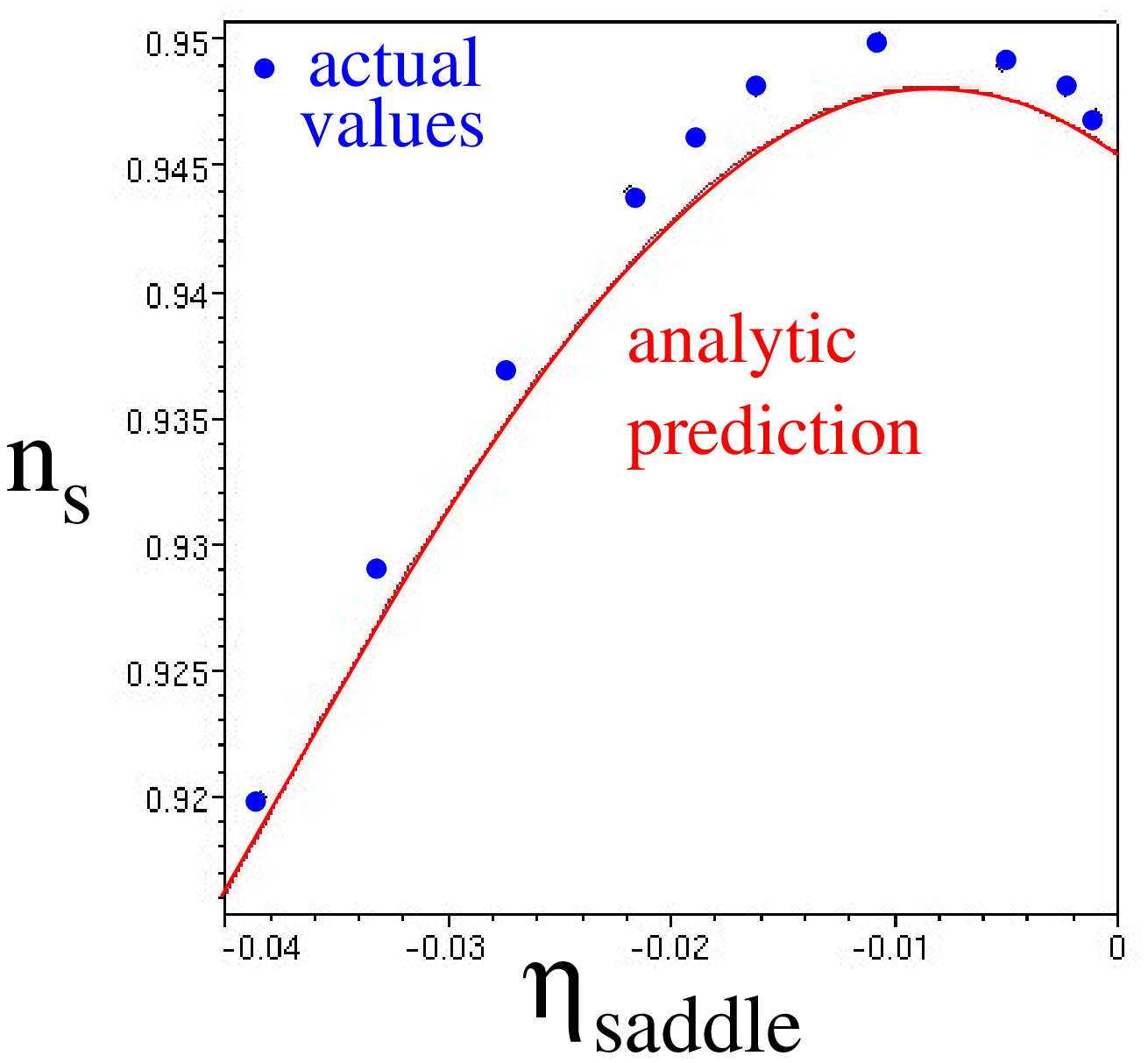,width=0.5\hsize}
\caption{Spectral index versus $\eta$ evaluated at the saddle
point; dots are exact numerical values, while the line is the
analytic prediction using the method of ref.\ \cite{postma}.  }
\label{nseta} }

Another potential source of tuning arises because the antipodal
point, $z = \frac12(1+\tau)$, where the inflationary saddle arises
is coincident with one of the $T_2/Z_2$ fixed points. As a result
one might worry about additional $D3$ interactions with the $D7$
and $O7$ planes that reside there. Although neither the
superpotential or uplifting physics is located at this point, this
need not preclude there being other $D3$-$D7$ instabilities which
might compete successfully with the inflationary slow roll. (On
the other hand, the same processes may be quite welcome once the
$D3$ encounters the $D7$ stack at the endpoint of the roll, when
$z = \frac12 \tau$.) To avoid these difficulties we therefore
demand either that no such physics exist (such as if the $D3$ and
relevant $D7$'s remain mutually BPS), or that the $D3$ not
approach the apex of the saddle point to within closer than the
string scale. This latter condition is easier to achieve the
larger is the torus volume, although there can be some tension
between having sufficiently large volumes and keeping the
potential large enough during inflation to get acceptably large
primordial scalar perturbations.

Not surprisingly, this scenario is very similar in its predictions
for primordial fluctuations to those of racetrack inflation
\cite{racetrack}, which is also based on axion motion from near a
saddle point in the potential. Just like in the racetrack model,
we find by numerical evolution that the spectral index --- which
we define at the canonical 55 $e$-foldings before the end of
inflation --- cannot exceed $n_s = 0.95$ even when the potential
is arbitrarily flat near the saddle point. A simple explanation of
the robustness of this result is given in ref.~\cite{postma},
which shows how it can be understood from the dominance of the
terms $V_0 - \frac12 m^2\phi^2$ in the inflaton potential until
the end of inflation. In figure \ref{nseta} we display the
variation of $n_s$ with the value of the $\eta$ parameter
evaluated at the saddle point ($\eta_{\rm saddle}$), both for the
exact numerical determination (dots) and the analytic
approximation (line) of ref.\ \cite{postma}.

%%%%%%%%%%%%%%%%%%%%%%%%%%%%%%%%%%%%%%%%%%%%%%%%%%%%%%%%%%%%%%%%%%%%%%%%

\subsection{A $D$-Term Driven Example}

Our second inflationary example is motivated by the inflationary
solution found in ref.~\cite{haack}, which we first briefly
describe.

\subsubsection*{The $D$-term inflationary setup}

Ref.~\cite{haack} seeks a stringy analogue of $D$-term inflation
\cite{DtermInf}. To do so they consider a nonperturbative
superpotential $W_{np}$ generated by a stack of $D7$ branes at the
fixed point $z = \frac12$ of $T_2/Z_2$. The inflationary energy
density and the uplifting potential is modelled as a
Fayet-Iliopoulos $D$-terms, given by adding fluxes to the $D7$
branes situated at $z = 0$. Inflation occurs when the $D3$ is in
close proximity to $z=0$, to which it falls driven by the one-loop
Coleman-Weinberg (CW) potential obtained using the threshold
corrections obtained from integrating out massive $D3$-$D7$ string
modes
\begin{equation}
 V_{FI} = \frac{g^2 \xi^2}{2} \left[ 1 + \frac{g^2 U(x)}{16\pi^2}
 \right] \,,
\end{equation}
where $g$ is the gauge coupling for the D7 brane at $z=0$, $\xi$
is a constant and $x$ is related to the canonically normalized
inflaton, $\phi \propto z_1$, by $x = \phi/\sqrt\xi$. (The large
mass associated with the $z_2$ coordinate allows it to be set
safely to zero.) The potential $U(x)$ is
\begin{equation}
 U(x) = (x^2 + 1)^2 \ln(x^2 + 1) + (x^2 - 1)^2 \ln (x^2 - 1) - 4\,
 x^4 \ln x - 4 \ln 2 \,.
\end{equation}
The slow roll occurs as $z$ rolls down the CW-potential, and ends
when the $D3$-$D7$ waterfall fields condense to cancel the
$D$-term. To obtain a Minkowski or dS vacuum after inflation an
additional uplifting term must be added, perhaps elsewhere on the
torus, although its explicit form is not considered in
ref.~\cite{haack}. This setup is illustrated in figure
\ref{d3d7fig}.

\FIGURE[ht]{ \epsfig{file=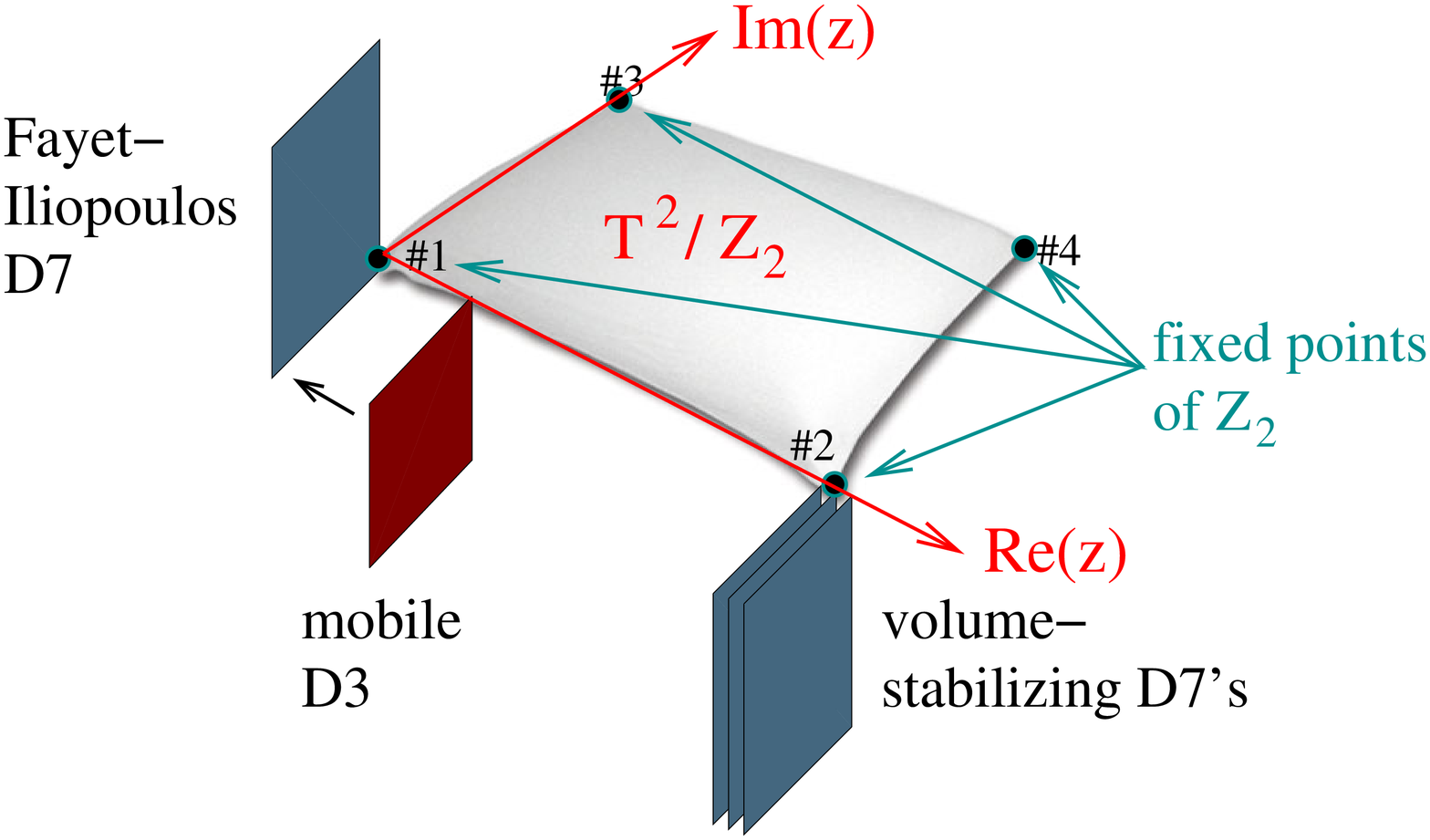,width=0.7\hsize} \caption{A
figure adapted from ref.\ \cite{haack} which illustrates their
inflationary configuration.} \label{d3d7fig} }

To gain an analytic understanding of the inflationary dynamics near
$\phi =0$ the authors of ref.\ \cite{haack} observe that $U(x)$ can be
well approximated by a logarithm in this regime~\footnote{The log
approximation is valid in the regime with $\phi \gg \xi$ during
inflation.}  , and the $F$-term potential can be expanded as a power
series about $z=0$, leading to
\begin{equation} \label{Vapprox1}
 V \simeq V_0 + D \ln
 \left( \frac{\phi^2}{\xi} \right) - \frac{m^2}{2} \,\phi^2
 + \frac{\lambda}{4} \, \phi^4 \,,
\end{equation}
where $V_0 = \frac12 \, g^2 \xi^2$ and $D = g^2 V_0/(8\pi^2)$. The
coefficients of this potential are regarded as implicit functions
of $S, T$ and $\tau$ --- all of which are regarded as being
stabilized. $m^2$ and $\lambda$ are obtained by expanding $V_F$ in
powers of $z$.
%, and the sign of the quadratic term is negative
%because the nonperturbative superpotential due to the brane at
%$z=\frac12$ tends to push the $D3$ away from $z=0$.

With these approximations the potential approaches that of usual
D-term hybrid inflation in the regime as $\phi \to 0$, where the
first two terms dominate. A slow roll in this regime is possible
provided $D/V_0 = g^2/(8\pi^2)$ can be made much smaller than
$(\phi/M_p)^2$. But this is not the only possibility; inflation
can also occur for larger $D3$-$D7$ separations. Indeed, provided
$m^4 > 4\lambda D$ the potential $V$ has a local maximum at
$\phi_{\rm max}^2 = (m^2 - \delta)/(2\lambda)$, whose curvature is
given by $V''(\phi_{\rm max}) = - 2 \delta$ where $\delta =
\sqrt{m^4 - 4\lambda D}$. A slow roll may therefore also be sought
near this local maximum. If $m^4 \simeq 4\lambda D$ (and so
$\delta \simeq 0$) this maximum coalesces with a local minimum at
$\phi^2_{\rm min} = (m^2 + \delta)/(2\lambda)$ to produce an
inflection point. The idea is thus to identify parameters to
ensure that the slow roll parameter $\eta \cong 0.015$, so that the
spectral index matches the WMAP5 value of $n_s \cong 1 - 2\eta =
0.96$, and once these are found see if such parameters can be
obtained from underlying brane dynamics on $K3\times T_2/Z_2$.

\subsubsection*{Supergravity search}

We seek to reproduce this scenario numerically within the
low-energy supergravity. This differs from the analysis of
ref.~\cite{haack} in two ways. First, we describe the $D$-term
physics using the full $D$-term potential, eq.~\pref{Dterm3fd},
when numerically seeking a slow roll. Second, we also add an
explicit uplifting term, which for definiteness we also take to be
a $D$-term arising from a flux localized on the $D7$'s located at
$z = \frac12$ ({\it i.e.,} at the same location as the gaugino
condensation $D7$'s). Using this potential we numerically compute
the potential for $z$ by evaluating the moduli fields at their
instantaneous minima $V(z) = V(z,S_0(z),T_0(z))$. We then search
for an inflationary slow roll with the $D3$ close to the $D7$'s at
$z=0$.

The best example of slow-roll inflation we found in this setup
arises near an inflection point of the scalar potential (described
below), corresponding to tuning the parameter $\delta \simeq 0$ in
the approximate potential of eq.~\pref{Vapprox1}. We did not find
examples of inflation arising at the local maximum, for which the
potential was also large enough to satisfy the COBE normalization.
We find that once $m^2$ is made sufficiently small to get a small
curvature at the maximum, the quartic term in eq.~\pref{Vapprox1}
becomes important, leading to the limit $\delta \simeq 0$. This
can be seen in the numerical evaluation of the potential shown in
figure \ref{fragile}.

Although we did examine a broad class of parameters, we were not
exhaustive enough to preclude the potential existence of the
inflationary example obtained in ref.~\cite{haack}. In particular,
our choice of $\mathfrak{f}_r = 0$ in eq.~\pref{frexpr} relates
our effective value for $g^2$ to the value of $s$ at its minimum,
and as a consequence the choices leading to a slow roll tend to
destabilize the potential for $s$. We do not know if the same need
be true once the freedom to adjust $\mathfrak{f}_r$ is used.

\FIGURE[ht]{ \centerline{\epsfig{file=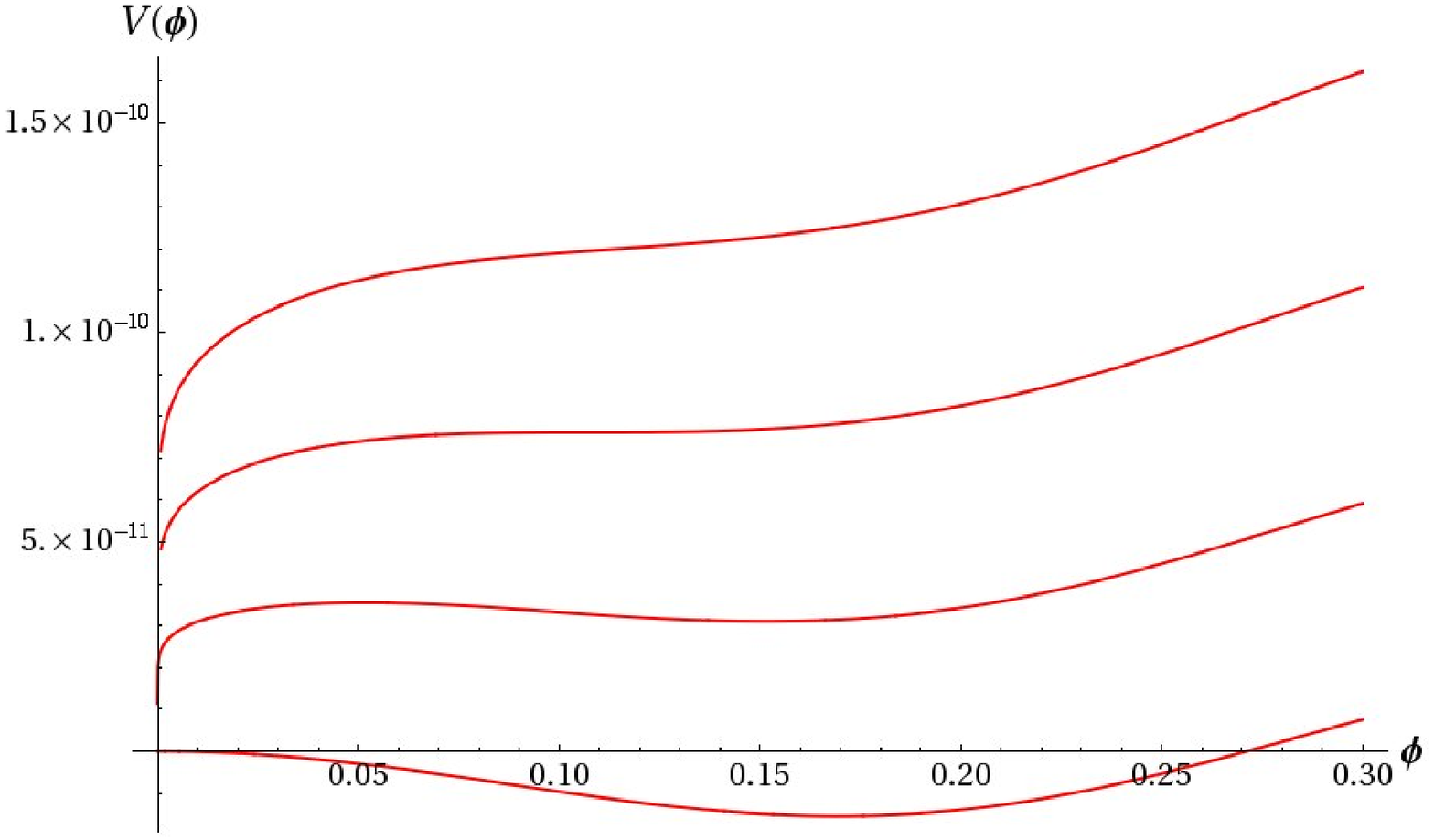,width=0.5\hsize}
\epsfig{file=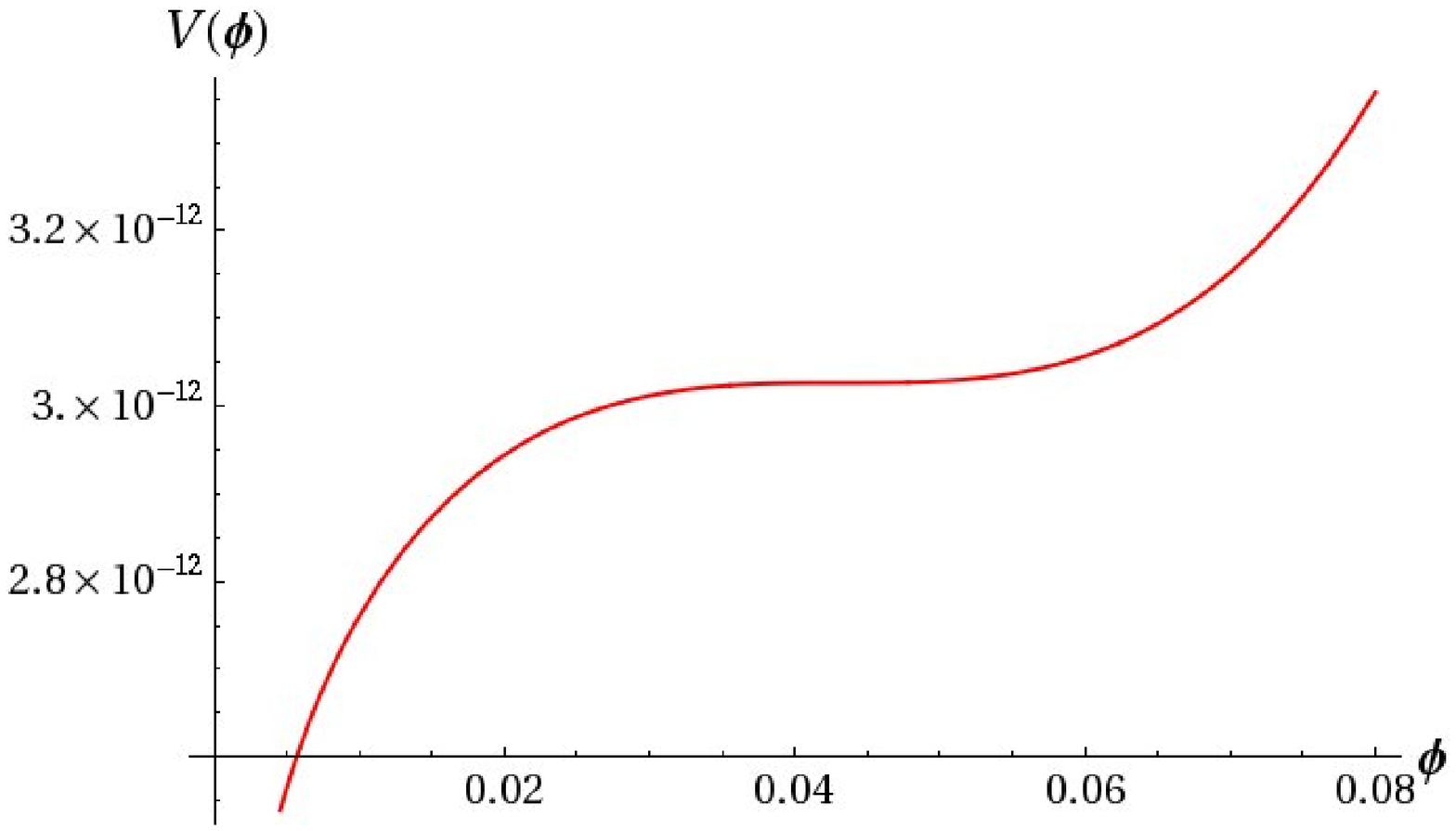,width=0.5\hsize}} \caption{Left:
Lowest curve is uplifted F-term potential; upper curves are $V_F +
\epsilon_i V_D$, where $V_D$ is the inflationary D-term, for an
increasing series of values of $\epsilon_i\approx 10^{-6}$ of the
strength which would be needed to uplift using this $V_D$ (from
$D7$'s at $z_1=0$) rather than the uplifting $V_D$ coming from
$D7$'s at $z_1=1/2$.  Right: fine-tuned inflection point potential
with parameters given by (\ref{inflect1},\ \ref{inflect2}).}
\label{fragile} }

Inflation at an inflection point can be found by tuning the
inflationary $D$-term (while adjusting the uplifting $D$-term to
ensure a Minkowski vacuum at the end of inflation). To see this,
choose for definiteness $\tau_1 = 0$ and $\tau_2 \lesssim 2$ and
turn off the inflationary $D$-term. Then the potential has a local
minimum at small values for $z_1$. As illustrated in figure
\ref{fragile}, this local minimum becomes increasingly shallow
until it eventually turns into the desired inflection point as the
inflationary $D$-term is turned back on.

A specific example uses $\tau_1 = 0$, $\tau_2 = 0.6$, $z_2 = 0$,
with $F$- and inflationary $D$-term potential parameters
\begin{equation}
 \quad N = 1,\ E_2 = 1.56512\times 10^{-8},\
 A_0 = \frac{1}{(2\pi)^{3/2}}
 \quad \hbox{and} \quad a= b=2\pi \,.
 \label{inflect1}
\end{equation}
where $E_2$ is the strength of the uplifting $D$-term located at
$z = \frac12$. The values of $B$ and $W_0$ are chosen by solving
$D_T W = D_S W = 0$, using eqs.~\pref{susysol}, to obtain minima
at $s_0 = t_0 = 5/\pi = 1.592$. Of these parameters only $\tau$ is
relatively important, since its value determines the sign of
$m^2$. The other parameters are randomly chosen, apart from the
coefficient of the inflationary D-term, whose value must be tuned
to
\be
    E_1 = 2.062673254\times 10^{-9}
\label{inflect2}
\ee
in order to obtain the desired inflection point, which occurs near
$z_1 = 0.107024$.

Starting sufficiently close to this point, and tuning $E_1$ as
above, one can obtain 5100 $e$-foldings of inflation. However,
this number decreases rapidly with less tuning. To get 60
$e$-foldings, $E_1$ must be increased only by 1 part in $10^6$,
otherwise the potential is not sufficiently flat. Other examples
we tried require comparable levels of fine-tuning, which is
somewhat more severe than the part-per-$10^3$ tuning that is
required in other brane-inflation models.

Although it is encouraging that inflation in this regime is
possible at all, it is a disadvantage that inflation occurs at an
inflection point rather than a maximum since this makes it is much
more sensitive to the initial conditions. For the above numbers
the initial value of $z_1$ cannot be increased by more than 6\%
from the inflection point without overshooting it and so ending
inflation too quickly. Furthermore, since inflation is not at a
maximum we cannot appeal to general arguments of eternal inflation
\cite{eternalinf} to explain these initial conditions.

\section{Conclusions}

This paper reports on the results of a detailed numerical search
for inflation in Type IIB vacua compactified (with
modulus-stabilizing fluxes) on $K3 \times T_2/Z_2$. The search is
performed using the 4D field equations of the low-energy effective
theory, which is constructed using familiar ingredients: an
$F$-term potential generated by fluxes and branes together with
some sort of uplifting physics.

There are two underlying motivations for performing this search.
The first starts from the observation that so much is known about
Type IIB compactifications on $K3 \times T_2/Z_2$, because much is
known about string behaviour on both $K3$ and the orbifolded
torus. The relative simplicity of these geometries makes the study
of their dynamics valuable, since most other instances of string
inflation arise in much more complicated contexts where
corrections can be more difficult to identify and control.

A second motivation for this study is the great appeal of the
$D3$-$D7$ inflationary mechanism \cite{d3d7}, which has the
promise of providing inflationary examples with a supersymmetric
final state, inflation driven by $D$ terms, and a potentially
interesting cosmic-string signature. $K3 \times T_2/Z_2$ is a
natural place to study this mechanism in detail because it
naturally contains stacks of $D7$ and $O7$ planes wrapping $K3$ as
well as a nice flat toroidal geometry in which to hope to find
slow-roll $D3$ motion. Furthermore, new tools to describe this
motion in terms of the low-energy 4D effective field theory have
recently been developed \cite{bhk,baumann,uplifting,haack}, and
their use allows a more systematic study of the degree to which
modulus stabilization interferes with the conditions required for
slow-roll inflation.

Using this 4D theory we numerically search for slow-roll
inflation. To do so we follow three of the possible low-energy
complex fields: the $D3$'s toroidal position, $z$; the $K3$ volume
modulus, $S$; and a modulus, $T$, dual to one of the 22 nontrivial
2-cycle volumes on $K3$. We consider two kinds of uplifting,
either that due to an anti-$D3$ brane \cite{KKLT}, or by a
flux-induced $D$ term potential \cite{bkq}. We follow earlier
workers in using this last type despite some of the consistency
problems \cite{HKLVZ,bkqplus} it raises when realized in string
vacua. We do so in the spirit that similar terms might arise from
more complicated string constructions, and it may therefore be
worth seeing whether they can support nontrivial inflationary
dynamics.

Our search identifies two kinds of slow-roll regime. What we regard to
be the most attractive has a $D3$ fall slowly between two $D7$ stacks,
driven by a modulus-stabilizing superpotential (perhaps produced by
gaugino condensation) located on a third stack. Uplifting is achieved
by adding a $\overline{D3}$ in a warped throat. A slow roll is then
possible when the brane is at the antipodal point from the
modulus-stabilizing stack, provided the torus is adjusted to be almost
perfectly square ({\it i.e.}  $\tau = i$) \footnote{Antipodal
inflation with unwarped $\overline{D3}$ lifting is also possible; in this case
the tuning of the torus is not so close to being square.}. The
low-energy scalar potential has a saddle point at this position, whose
unstable inflaton direction turns out to be a linear combination of
the $D3$ position, $z_1$, and the axion, $\alpha = \hbox{Im} \, S$,
associated with the $K3$ volume modulus. The resulting inflationary
picture resembles earlier `racetrack' models \cite{racetrack}, and
shares their generic prediction $n_s \lsim 0.95$. Although the
inflation is robust against changes to the superpotential parameters,
it is sensitive to the kind of uplifting involved and requires a
1-in-$10^4$ tuning in the value of $\tau$.

The second inflationary regime found generates a superpotential
(and places an uplifting flux) at one fixed point, $z=\frac12$,
and places another inflationary $D$-term generating flux on a
second brane stack at $z=0$. Inflation is then sought with the
$D3$ very close to $z=0$, in the hopes of obtaining standard
hybrid $D$-term inflation as the $D3$ dissolves into the $D7$'s
there. Our search here led to inflation at an inflection point,
provided the inflationary $D$ term is tuned to a part in $10^6$.
But because it arises at an inflection point, this inflationary
scenario is sensitive to the inflaton's initial conditions due to
a potential overshoot problem. In this case use of the full
effective 4D potential makes finding inflation more difficult than
might be thought based on simpler approximate potentials.

\section*{Acknowledgments}

We wish to thank Keshav Dasgupta, Michael Haack, Renata Kallosh,
Axel Krause, Andre Linde, Dieter L\"ust, Liam McAllister, Fernando
Quevedo and Marco Zagermann for many helpful comments and
suggestions. CB and JC also wish to thank the Banff International
Research Station for providing the lovely setting where this work
was begun, as well as the Natural Sciences and Engineering
Research Council (NSERC) of Canada whose funds provide partial
research support. MP is supported by a VIDI grant from the Dutch
Organisation for Scientific Research (NWO). Research at the
Perimeter Institute is supported in part by the Government of
Canada through NSERC and by the Province of Ontario through MRI.

\appendix

%%%%%%%%%%%%%%%%%%%%%%%%%%%%%%%%%%%%%%%%%%%%%%%%%%%%%%%%%%%%%%%%%%%%%%%%%%%
%%%%%%%%%%%%%%%%%%%%%%%%%%%%%%%%%%%%%%%%%%%%%%%%%%%%%%%%%%%%%%%%%%%%%%%%%%%
%%%%%%%%%%%%%%%%%%%%%%%%%%%%%%%%%%%%%%%%%%%%%%%%%%%%%%%%%%%%%%%%%%%%%%%%%
\section{Useful relations}

In this appendix we collect various useful relations and
identities that are used in the text.

\subsection{No-Scale Condition}
\label{App:Noscale}

We briefly show here that any K\"ahler function, $K(T,\overline T)
= K(T+\overline T)$, which satisfies the scaling identity
\be \label{app:scaling}
    K(\lambda T, \lambda \overline T) \equiv
    K(T, \overline T) - 3\ln \lambda \,,
\ee
for arbitrary moduli $T^\alpha$ and constant $\lambda$, must also
satisfy the no-scale condition
\be \label{app:noscale}
    K^{\alpha\overline\beta} K_\alpha K_{\overline \beta}
    \equiv 3 \,.
\ee
This establishes the no-scale property of the K\"ahler function of
interest in the main text, $K = K(S+\overline S,T_i + \overline
T_i,i(z-\bar z),i(\tau - \overline \tau))$, which satisfies these
assumptions.

To establish the result we first recognize that because $K$ is a
real function only of the combination $(T+\overline T)^\alpha$, we
may ignore the distinction between derivatives with respect to
$T^\alpha$ and $\overline T^{\overline \alpha}$: $K_\alpha =
K_{\overline \alpha} = \partial K/\partial X^\alpha$, where
$X^\alpha = (T +\overline T)^\alpha$. Next we differentiate
eq.\ \pref{app:scaling} once with respect to $\lambda$, and then a
second time with respect to $T^\alpha$, giving
\ba \label{app:dlambda}
    X^\beta K_\beta(\lambda X)
    &\equiv&
    - \frac{3}{\lambda} \qquad\quad (\partial/\partial \lambda )\\
    K_\alpha(\lambda X)
    + \lambda X^\beta K_{\alpha\beta}(\lambda X)
    &\equiv& 0 \qquad\qquad
    (\partial^2/\partial T^\alpha \partial\lambda) \,.
    \label{app:dlambdadX}
\ea
Contracting two copies of eq.\ \pref{app:dlambdadX} together using
the inverse matrix $K^{\alpha\gamma}$ then gives
\ba
    K^{\alpha\gamma}(\lambda X) K_\alpha (\lambda X)
    K_\gamma (\lambda X) &=& \lambda^2
    K^{\alpha \gamma}(\lambda X) K_{\alpha \xi}(\lambda X)
    K_{\gamma \rho}(\lambda X) X^\xi X^\rho \nn\\
    &=& \lambda^2 K_{\alpha \beta} (\lambda X)
    X^\alpha X^\beta \,,
\ea
which may be further simplified by contracting
eq.\ \pref{app:dlambdadX} with $X^\alpha$ and using
eq.\ \pref{app:dlambda}, to get
\be
    K^{\alpha\gamma}(\lambda X) K_\alpha (\lambda X)
    K_\gamma (\lambda X) = \lambda^2 K_{\alpha \beta} (\lambda X)
    X^\alpha X^\beta = 3 \,.
\ee
The desired result is now obtained by evaluating at $\lambda = 1$.

%%%%%%%%%%%%%%%%%%%%%%%%%%%%%%%%%%%%%%%%%%%%%%%%%%%%%%%%%%%%%%%%%%%%%%%%%%
\subsection{Theta Functions and Periodicity}
\label{s:periodicity} \label{s:theta}

We adopt the following definition for the Jacobi theta function
\ba
    \vartheta_1(u|\tau) = \vartheta_1(u;q)
    &\equiv& -i\sum_{n=-\infty}^\infty (-)^n e^{(2n+1) i u}
    q^{(n+1/2)^2} \nn\\
    &=& 2 q^{1/4} \sum_{n=0}^\infty (-)^n q^{n(n+1)}
    \sin[(2n+1) u] \,,
\ea
with $q = e^{i\pi \tau}$. This satisfies
\ba
    \vartheta_1(u \pm \pi|\tau) &=& - \vartheta_1(u|\tau) \nn\\
    \vartheta_1(u \pm \pi \tau|\tau) &=& - q^{-1} e^{\mp 2 i u}
    \vartheta_1(u|\tau) \,,
\ea
under the displacements that define the periods of the torus
$T_2$.

The combination $F_r(z,\tau) = \vartheta_1[\pi(z_r - z)|\tau]
\vartheta_1[\pi(z_r + z)|\tau]$ appearing in the superpotential
therefore transforms as
\ba
    F_r(z+1,\tau) &=& F_r(z,\tau) \nn\\
    F_r(z+\tau,\tau) &=& e^{-4i\pi z - 2i\pi \tau} F_r(z, \tau) \,,
\ea
for any $z_r$. Clearly the combination $e^{-aS} F_r(z,\tau)$ is
therefore invariant under the combined transformations $(z,S) \to
(z+1,S)$ and
\be
    z \to z+\tau \,, \qquad
    S \to - \frac{2\pi i}{a}(2z +\tau) \,.
\ee

When $\tau_2 \gsim 1$ we have $|q| \ll 1$ and so the above series
for $\vartheta_1$ is well approximated by its first terms,
$\vartheta_1(u|\tau) \simeq 2 q^{1/4} \sin u$, and so
\be
    F_r \simeq 4 q^{1/2} \sin [ \pi(z_r - z) ] \sin [\pi (z_r + z)]
    = 2 q^{1/2} \Bigl[ \cos(2\pi z) - \cos(2 \pi z_r) \Bigr]
    \,.
\ee
Using $z_0 = 0$, $z_1 = \frac12$, $z_2 = \frac12 \tau$ and $z_3 =
\frac12(1+\tau)$, we have respectively $\cos(2\pi z_0) = 1$,
$\cos(2\pi z_1) = -1$, $\cos(2\pi z_2) = \cos(\pi\tau)$ and
$\cos(2\pi z_3) = -\cos(\pi\tau)$. Notice in particular that in
this limit
\be
    \sum_{r=0,1} F_r \simeq \sum_{r=2,3} F_r
    \simeq \frac12 \sum_{r=0}^3 F_r
    \simeq4  q^{1/2} \cos(2\pi z) \,.
\label{largetau} \ee

%%%%%%%%%%%%%%%%%%%%%%%%%%%%%%%%%%%%%%%%%%%%%%%%%%%%%%%%%%%%%%%%%%%%%%%%%%
\subsection{Scaling behavior}
\label{s:scaling}

This appendix displays a useful scaling property that allows one
to relate numerical results for different choices of parameters.

The scalar potential arises as the sum of an $F$-term, $D$-term
and an up-lifting contribution, $V = V_F + V_D + V_{\rm up}$, with
the $D$-term and uplifting contributions having the form
\be
    V_{\rm up} = \frac{E}{X^m t^n} \,,
\ee
and
\be
    V_D = \frac{E_r}{\hbox{Re} f_r \, t^2} \,,
\ee
where $t = \hbox{Re}\, T$, $E$ and $E_r$ are constants, $X = 2[s -
c \, g(z,\bar z)]$, $s = \hbox{Re}\, S$ and $f_r = S - (1/a)
h(z)$. Here $g$ and $h$ are functions whose form is not important
in what follows. The constant $c$ is related to $a$ by $ac =
2\pi$, due to the requirement that the potential be periodic under
the toroidal shift $z \to z + \tau$. The $F$-term potential is
similarly computed using the K\"ahler potential
\be
 K = - \ln X - 2\ln( T + \overline T) \,,
\ee
and superpotential
\be
 W = W_0 + A(z) e^{-aS} + B e^{-bT} \,.
\ee
Here $W_0$, $B$, $b$ and $a$ are constants --- with $a$ the same
constant as appears in $f_r$ --- and $A(z)$ is a function whose
detailed form is not important for the argument now to be made.

These contributions to the scalar potential have the property that
they scale simply under the following redefinitions
\be
    s \to \lambda_1 s, \qquad t \to \lambda_2 t, \qquad a
    \to a/\lambda_1 , \qquad b \to b/\lambda_2 , \quad
    \hbox{and} \quad E \to
    \lambda_1^{m-1} \lambda_2^{n-2} \,,\label{scaling1}
\ee
with $z$, $E_r$, $A$, $B$ and $W_0$ held fixed. With these choices
we have $X \to \lambda_1 X$, $\hbox{Re}f_r \to \lambda_1
\hbox{Re}f_r$ and so $W \to W$, $e^K \to e^K/(\lambda_1
\lambda_2^2)$. This makes $V_F$, $V_D$ and $V_{\rm up}$ all scale
very simply:
\be
 V  \to \frac{V}{\lambda_1 \lambda_2^2} \,.
\ee

Because $V_F$ is quadratic in $W$ and its derivatives, the
$F$-term potential also rescales as $V_F \to \lambda^2 V_F$ under
the scalings
\be
    x \to \lambda x, \qquad \hbox{where}
    \qquad x =\{A,B,W_0\}
    \label{scaling2} \,.
\ee

%%%%%%%%%%%%%%%%%%%%%%%%%%%%%%%%%%%%%%%%%%%%%%%%%%%%%%%%%%%%%%%%%%%%%%%%%%%
%%%%%%%%%%%%%%%%%%%%%%%%%%%%%%%%%%%%%%%%%%%%%%%%%%%%%%%%%%%%%%%%%%%%%%%%%%%
%%%%%%%%%%%%%%%%%%%%%%%%%%%%%%%%%%%%%%%%%%%%%%%%%%%%%%%%%%%%%%%%%%%%%%%%%%%%


\begin{thebibliography}{9}

\bibitem{revs}
For recent reviews with references see
  F.~Quevedo,
  %``Lectures on string/brane cosmology,''
  Class.\ Quant.\ Grav.\  {\bf 19} (2002) 5721
  [hep-th/0210292];
  %%CITATION = CQGRD,19,5721;%%
  %
    A.~Linde,
  ``Inflation and string cosmology,''
  eConf {\bf C040802} (2004) L024
  [J.\ Phys.\ Conf.\ Ser.\  {\bf 24} (2005) 151]
  [hep-th/0503195];
  %%CITATION = HEP-TH 0503195;%%
%
    %
    S.~H.~Henry Tye,
  %``Brane inflation: String theory viewed from the cosmos,''
  [hep-th/0610221];
  %%CITATION = HEP-TH/0610221;%%
    %
  J.~M.~Cline,
  ``String cosmology,''
  [hep-th/0612129];
  %%CITATION = HEP-TH/0612129;%%
  %
C.~P.~Burgess,
  %``Lectures on Cosmic Inflation and its Potential Stringy Realizations,''
  PoS {\bf P2GC} (2006) 008
  [Class.\ Quant.\ Grav.\  {\bf 24} (2007) S795]
  [arXiv:0708.2865 [hep-th]];
  %%CITATION = CQGRD,24,S795;%%
%
R.~Kallosh,
  %``On Inflation in String Theory,''
  Lect.\ Notes Phys.\  {\bf 738} (2008) 119
  [hep-th/0702059];
  %%CITATION = LNPHA,738,119;%%
  %
L.~McAllister and E.~Silverstein,
  %``String Cosmology: A Review,''
  Gen.\ Rel.\ Grav.\  {\bf 40} (2008) 565
  [arXiv:0710.2951 [hep-th]].
  %%CITATION = GRGVA,40,565;%%

\bibitem{d3d7}
 C.~Herdeiro, S.~Hirano and R.~Kallosh,
  %``String theory and hybrid inflation / acceleration,''
  JHEP {\bf 0112} (2001) 027
  [hep-th/0110271];
  %%CITATION = JHEPA,0112,027;%%
  %
K.~Dasgupta, C.~Herdeiro, S.~Hirano and R.~Kallosh,
%``D3/D7 inflationary model and M-theory,''
  Phys.\ Rev.\ D {\bf 65}, 126002 (2002)
  [hep-th/0203019];
  %%CITATION = HEP-TH 0203019;%%
  %
J.~P.~Hsu, R.~Kallosh and S.~Prokushkin,
%``On brane inflation with volume stabilization,''
JCAP {\bf 0312} (2003) 009 [hep-th/0311077];
%%CITATION = HEP-TH 0311077;%%
%
%\bibitem{Koyama:2003yc}
F.~Koyama, Y.~Tachikawa and T.~Watari,
%``Supergravity analysis of hybrid inflation
%model from D3-D7 system'',
[hep-th/0311191];
%%CITATION = HEP-TH 0311191;%%
%
J.~P.~Hsu and R.~Kallosh,
%``Volume stabilization and the origin of the
%inflaton shift symmetry in string theory,''
JHEP {\bf 0404} (2004) 042 [hep-th/0402047].
%%CITATION = HEP-TH 0402047;%%
%
K.~Dasgupta, J.~P.~Hsu, R.~Kallosh, A.~Linde and M.~Zagermann,
  %``D3/D7 brane inflation and semilocal strings,''
  JHEP {\bf 0408}, 030 (2004)
  [hep-th/0405247];
  %%CITATION = HEP-TH 0405247;%%
%
P.~Chen, K.~Dasgupta, K.~Narayan, M.~Shmakova and M.~Zagermann,
  %``Brane inflation, solitons and cosmological solutions: I,''
  JHEP {\bf 0509}, 009 (2005)
  [hep-th/0501185];
  %%CITATION = HEP-TH 0501185;%%
  %
  L.~McAllister,
  %``An inflaton mass problem in string
  %inflation from threshold corrections  to
  %volume stabilization,''
  JCAP {\bf 0602} (2006) 010
  [hep-th/0502001].
  %%CITATION = JCAPA,0602,010;%%

\bibitem{DT}
G.~R.~Dvali and S.~H.~H.~Tye,
%``Brane inflation,''
Phys.\ Lett.\ B {\bf 450} (1999) 72 [hep-ph/9812483].
%%CITATION = HEP-PH 9812483;%%

\bibitem{BBar}
C.~P.~Burgess, M.~Majumdar, D.~Nolte, F.~Quevedo, G.~Rajesh and
R.~J.~Zhang,
%``The inflationary brane-antibrane universe,''
JHEP {\bf 0107} (2001) 047 [hep-th/0105204].
%%CITATION = HEP-TH 0105204;%%

\bibitem{BBar2}
G.~R.~Dvali, Q.~Shafi and S.~Solganik,
%``D-brane inflation,''
hep-th/0105203.
%%CITATION = HEP-TH 0105203;%%

\bibitem{KKLMMT}
S.~Kachru, R.~Kallosh, A.~Linde, J.~Maldacena, L.~McAllister and
S.~P.~Trivedi, %``Towards inflation in string theory,''
JCAP {\bf 0310} (2003) 013 [hep-th/0308055];
  %%CITATION = HEP-TH 0308055;%%
%
C.~P.~Burgess, J.~M.~Cline, H.~Stoica and F.~Quevedo,
%``Inflation in realistic D-brane models,''
JHEP {\bf 0409} (2004) 033 [hep-th/0403119].
%%CITATION = HEP-TH 0403119;%%

\bibitem{GKP}
S.~B.~Giddings, S.~Kachru and J.~Polchinski, %``Hierarchies from
%fluxes in string compactifications,''
Phys. Rev. {\bf D66}, 106006 (2002);
%
S.~Sethi, C.~Vafa and E.~Witten, %``Constraints on low-dimensional
%string compactifications,''
Nucl.\ Phys.\ B {\bf 480} (1996) 213 [hep-th/9606122].

\bibitem{dasgupta}
K.~Dasgupta, G.~Rajesh and S.~Sethi,
  %``M theory, orientifolds and G-flux,''
  JHEP {\bf 9908} (1999) 023
  [arXiv:hep-th/9908088].
  %%CITATION = JHEPA,9908,023;%%

\bibitem{KKLT}
S. Kachru, R. Kallosh, A. Linde and S. P. Trivedi, %``de Sitter
%Vacua in String Theory,''
Phys.\ Rev.\ D {\bf 68} (2003) 046005 [ hep-th/0301240];
%
 %
 B.~S.~Acharya,
  %``A moduli fixing mechanism in M theory,''
  [hep-th/0212294];
  %%CITATION = HEP-TH 0212294;%%
  %
   R.~Brustein and S.~P.~de Alwis,
  %``Moduli potentials in string compactifications with fluxes: Mapping the
  %discretuum,''
  Phys.\ Rev.\ D {\bf 69} (2004) 126006
  [hep-th/0402088];
  %%CITATION = HEP-TH 0402088;%%
  %
   F.~Denef, M.~R.~Douglas, B.~Florea, A.~Grassi and S.~Kachru,
  %``Fixing all moduli in a simple F-theory compactification,''
  [hep-th/0503124].
  %%CITATION = HEP-TH 0503124;%%

\bibitem{tt}
P.~K.~Tripathy and S.~P.~Trivedi,
  %``Compactification with flux on K3 and tori,''
  JHEP {\bf 0303} (2003) 028
  [arXiv:hep-th/0301139].
  %%CITATION = JHEPA,0303,028;%%

\bibitem{ak}
P.~S.~Aspinwall and R.~Kallosh,
  %``Fixing all moduli for M-theory on K3 x K3,''
  JHEP {\bf 0510} (2005) 001
  [arXiv:hep-th/0506014].
  %%CITATION = JHEPA,0510,001;%%

\bibitem{GKTT}
L. G\"orlich, S. Kachru, P.K. Tripathy and S.P. Trivedi,
  %``Gaugino Condensation and Nonperturbative Superpotentials
  %in Flux Compactifications,''
  JHEP {\bf 0412} (2004) 074
  [arXiv:hep-th/0407130].
  %%CITATION = JHEPA,0412,074;%%

\bibitem{bhk}
 M.~Berg, M.~Haack and B.~Kors,
  %``Loop corrections to volume moduli and inflation in string theory,''
  Phys.\ Rev.\  D {\bf 71} (2005) 026005
  [hep-th/0404087];
  %%CITATION = PHRVA,D71,026005;%%
%
  M.~Berg, M.~Haack and B.~Kors,
  %``String loop corrections to Kaehler potentials in orientifolds,''
  JHEP {\bf 0511} (2005) 030
  [hep-th/0508043].
  %%CITATION = JHEPA,0511,030;%%

\bibitem{baumann}
D.~Baumann, A.~Dymarsky, I.~R.~Klebanov, J.~M.~Maldacena,
L.~P.~McAllister and A.~Murugan,
  %``On D3-brane potentials in compactifications with fluxes and wrapped
  %D-branes,''
  JHEP {\bf 0611} (2006) 031
  [arXiv:hep-th/0607050].
  %%CITATION = JHEPA,0611,031;%%

\bibitem{uplifting}
C.~P.~Burgess, J.~M.~Cline, K.~Dasgupta and H.~Firouzjahi,
  %``Uplifting and inflation with D3 branes,''
  JHEP {\bf 0703} (2007) 027
  [arXiv:hep-th/0610320].
  %%CITATION = JHEPA,0703,027;%%

\bibitem{haack}
M.~Haack, R.~Kallosh, A.~Krause, A.~Linde, D.~Lust and
M.~Zagermann,
  %``Update of D3/D7-Brane Inflation on K3 x T^2/Z_2,''
  arXiv:0804.3961 [hep-th].
  %%CITATION = ARXIV:0804.3961;%%

\bibitem{bkq}
C.~P.~Burgess, R.~Kallosh and F.~Quevedo, %``de Sitter string vacua
%from supersymmetric D-terms,''
JHEP 0310 (2003) 056, [hep-th/0309187].
%%CITATION = HEP-TH 0309187;%%

\bibitem{racetrack}
  J.~J.~Blanco-Pillado {\it et al.},
  %``Racetrack inflation,''
  JHEP {\bf 0411}, 063 (2004)
  [arXiv:hep-th/0406230];
  %%CITATION = JHEPA,0411,063;%%
  %``Inflating in a better racetrack,''
  JHEP {\bf 0609}, 002 (2006)
  [arXiv:hep-th/0603129].
  %%CITATION = JHEPA,0609,002;%%

\bibitem{strominger}
A.~Strominger,
  %``Superstrings with Torsion,''
  Nucl.\ Phys.\  B {\bf 274} (1986) 253.
  %%CITATION = NUPHA,B274,253;%%

\bibitem{K3}
P.~S.~Aspinwall,
  %``K3 surfaces and string duality,''
  arXiv:hep-th/9611137.
  %%CITATION = HEP-TH/9611137;%%


\bibitem{HKLVZ}
  M.~Haack, D.~Krefl, D.~Lust, A.~Van Proeyen and M.~Zagermann,
  %``Gaugino condensates and D-terms from D7-branes,''
  JHEP {\bf 0701} (2007) 078
  [arXiv:hep-th/0609211].
  %%CITATION = JHEPA,0701,078;%%

\bibitem{bkqplus}
K. Choi, A. Falkowski, H.P. Nilles and M. Olechowski,
%``Soft supersymmetry breaking in KKLT flux compactification,''
Nucl.\ Phys.\ {\bf B718} (2005) 113 [hep-th/0503216];
%
S.P. de Alwis,
%``Effective potentials for light moduli,''
Phys.\ Lett.\ {\bf B626} (2005) 223 [hep-th/0506266];
%
 G.~Villadoro and F.~Zwirner,
  %``de Sitter vacua via consistent D-terms,''
  Phys.\ Rev.\ Lett.\  {\bf 95} (2005) 231602
  [hep-th/0508167];
  %%CITATION = HEP-TH 0508167;%%
%
A. Achucarro, B. de Carlos, J.A. Casas and L. Doplicher,
%``de Sitter vacua from uplifting D-terms in
%effective supergravities from realistic strings,''
JHEP {\bf 0606}, 014 (2006) [hep-th/0601190];
%
  G.~Villadoro and F.~Zwirner,
   %``D terms from D-branes, gauge invariance and
   %moduli stabilization in flux compactifications,''
  JHEP {\bf 0603} (2006) 087
  [hep-th/0602120];
  %%CITATION = HEP-TH 0602120;%%
  %
  Ph.~Brax, C.~.~v.~de Bruck, A.~C.~Davis,
  S.~C.~Davis, R.~Jeannerot and M.~Postma,
  %``Moduli corrections to D-term inflation,''
  [hep-th/0610195];
  %%CITATION = HEP-TH 0610195;%%
%
 D.~Cremades, M.~P.~Garcia del Moral, F.~Quevedo and K.~Suruliz,
  %``Moduli stabilisation and de Sitter string vacua from magnetised D7
  %branes,''
  JHEP {\bf 0705} (2007) 100
  [arXiv:hep-th/0701154];
  %%CITATION = JHEPA,0705,100;%%
%
 B.~de Carlos, J.~A.~Casas, A.~Guarino, J.~M.~Moreno and O.~Seto,
  %``Inflation in uplifted supergravities,''
  JCAP {\bf 0705} (2007) 002
  [arXiv:hep-th/0702103];
  %%CITATION = JCAPA,0705,002;%%
%
  F.~Chen and H.~Firouzjahi,
  %``Dynamics of D3-D7 Brane Inflation in Throats,''
  arXiv:0807.2817 [hep-th].
  %%CITATION = ARXIV:0807.2817;%%

\bibitem{jockers}
M.~Grana, T.~W.~Grimm, H.~Jockers and J.~Louis,
  %``Soft supersymmetry breaking in Calabi-Yau orientifolds with D-branes  and
  %fluxes,''
  Nucl.\ Phys.\  B {\bf 690} (2004) 21
  [arXiv:hep-th/0312232];
  %%CITATION = NUPHA,B690,21;%%
%
H.~Jockers and J.~Louis,
  %``The effective action of D7-branes in N = 1 Calabi-Yau orientifolds,''
  Nucl.\ Phys.\  B {\bf 705} (2005) 167
  [arXiv:hep-th/0409098];
  %%CITATION = NUPHA,B705,167;%%
%
H.~Jockers and J.~Louis,
  %``D-terms and F-terms from D7-brane fluxes,''
  Nucl.\ Phys.\  B {\bf 718} (2005) 203
  [arXiv:hep-th/0502059].
  %%CITATION = NUPHA,B718,203;%%

\bibitem{warpSUSY}
   O.~DeWolfe and S.~B.~Giddings,
  %``Scales and hierarchies in warped compactifications and brane worlds,''
  Phys.\ Rev.\  D {\bf 67} (2003) 066008
  [arXiv:hep-th/0208123];
  %%CITATION = PHRVA,D67,066008;%%
%
  S.~B.~Giddings and A.~Maharana,
  %``Dynamics of warped compactifications and the shape of the warped
  %landscape,''
  Phys.\ Rev.\  D {\bf 73} (2006) 126003
  [arXiv:hep-th/0507158];
  %%CITATION = PHRVA,D73,126003;%%
  %
  C.~P.~Burgess, P.~G.~Camara, S.~P.~de Alwis, S.~B.~Giddings,
  A.~Maharana, F.~Quevedo and K.~Suruliz,
  %``Warped supersymmetry breaking,''
  JHEP {\bf 0804} (2008) 053
  [arXiv:hep-th/0610255],
  %%CITATION = JHEPA,0804,053;%%
%
    G.~Shiu, G.~Torroba, B.~Underwood and M.~R.~Douglas,
  %``Dynamics of Warped Flux Compactifications,''
  JHEP {\bf 0806} (2008) 024
  [arXiv:0803.3068 [hep-th]],
  %%CITATION = JHEPA,0806,024;%%
%
  A.~R.~Frey, G.~Torroba, B.~Underwood and M.~R.~Douglas,
  %``The Universal Kaehler Modulus in Warped Compactifications,''
  arXiv:0810.5768 [hep-th].
  %%CITATION = ARXIV:0810.5768;%%


\bibitem{baumann2}
  D.~Baumann, A.~Dymarsky, I.~R.~Klebanov, L.~McAllister and P.~J.~Steinhardt,
  %``A Delicate Universe,''
  Phys.\ Rev.\ Lett.\  {\bf 99}, 141601 (2007)
  [arXiv:0705.3837 [hep-th]];
  %%CITATION = PRLTA,99,141601;%%
  %
 D.~Baumann, A.~Dymarsky, I.~R.~Klebanov and L.~McAllister,
  %``Towards an Explicit Model of D-brane Inflation,''
  JCAP {\bf 0801}, 024 (2008)
  [arXiv:0706.0360 [hep-th]].
  %%CITATION = JCAPA,0801,024;%%

\bibitem{KMI}
J.~P.~Conlon and F.~Quevedo,
  %``K\"ahler moduli inflation,''
  JHEP {\bf 0601} (2006) 146
  [hep-th/0509012];
  %%CITATION = HEP-TH 0509012;%%
%
 J.~Simon, R.~Jimenez, L.~Verde, P.~Berglund and V.~Balasubramanian,
  %``Using cosmology to constrain the topology of hidden dimensions,''
  [astro-ph/0605371];
  %%CITATION = ASTRO-PH 0605371;%%
  %
 J.~R.~Bond, L.~Kofman, S.~Prokushkin and P.~M.~Vaudrevange,
  %``Roulette inflation with Kaehler moduli and their axions,''
  Phys.\ Rev.\  D {\bf 75} (2007) 123511
  [arXiv:hep-th/0612197];
  %%CITATION = PHRVA,D75,123511;%%
%
M.~Cicoli, C.~P.~Burgess and F.~Quevedo,
  %``Fibre Inflation: Observable Gravity Waves from IIB String
  %Compactifications,''
  [arXiv:0808.0691 [hep-th]].
  %%CITATION = ARXIV:0808.0691;%%

\bibitem{BBHL}
K.~Becker, M.~Becker, M.~Haack and J.~Louis,
  %``Supersymmetry breaking and alpha'-corrections to flux induced
  % potentials,''
  JHEP {\bf 0206} (2002) 060
  [hep-th/0204254].
  %%CITATION = JHEPA,0206,060;%%

 \bibitem{eternalinf}
   A.~D.~Linde,
  %``Monopoles as big as a universe,''
  Phys.\ Lett.\ B {\bf 327}, 208 (1994)
  [arXiv:astro-ph/9402031];
  %%CITATION = ASTRO-PH 9402031;%%
%
  A.~Vilenkin,
  %``Topological inflation,''
  Phys.\ Rev.\ Lett.\  {\bf 72}, 3137 (1994)
  [arXiv:hep-th/9402085].
  %%CITATION = HEP-TH 9402085;%%

\bibitem{postma}
  Ph.~Brax, S.~C.~Davis and M.~Postma,
  ``The Robustness of $n_s < 0.95$ in Racetrack Inflation,''
  JCAP {\bf 0802}, 020 (2008)
  [arXiv:0712.0535 [hep-th]].
  %%CITATION = JCAPA,0802,020;%%

\bibitem{DtermInf}
 P. Binetruy and G. R. Dvali,
 %``D-term inflation,''
 Phys.\ Lett.\ {\bf B 388} (1996) 241
 [arXiv:hep-ph/9606342];
%
 E. Halyo,
 %``Hybrid inflation from supergravity D-terms,''
 Phys.\ Lett.\ {\bf B 387} (1996) 43
 [arXiv:hep-ph/9606423].


\end{thebibliography}
\end{document}